\documentclass[12pt]{article}
\usepackage{a4wide}
\usepackage{amssymb}
\usepackage{amsmath}
\usepackage{graphicx}
\begin{document}
{\renewcommand{\thefootnote}{\fnsymbol{footnote}}
\hfill  AEI--2005--169, NI05063\\
\medskip
\hfill math-ph/0511043\\
\medskip
\begin{center}
{\LARGE  Effective Equations of Motion for Quantum Systems}\\
\vspace{1.5em}
Martin Bojowald\footnote{e-mail address: {\tt mabo@aei.mpg.de}}
and  Aureliano Skirzewski\footnote{e-mail address: {\tt skirz@aei.mpg.de}}
\\
\vspace{0.5em}
Max-Planck-Institut f\"ur Gravitationsphysik, Albert-Einstein-Institut,\\
Am M\"uhlenberg 1, D-14476 Potsdam, Germany
\vspace{1.5em}
\end{center}
}

\setcounter{footnote}{0}

\newtheorem{theo}{Theorem}
\newtheorem{lemma}{Lemma}
\newtheorem{corr}{Corollary}
\newtheorem{defi}{Definition}

\newcommand{\proofend}{\raisebox{1.3mm}{\fbox{\begin{minipage}[b][0cm][b]{0cm}
\end{minipage}}}}
\newenvironment{proof}{\noindent{\it Proof:} }{\mbox{}\hfill
\proofend\\\mbox{}}

\newcounter{exam}
\newenvironment{excount}[1]{\smallskip\stepcounter{exam}\noindent{\bf Example %
\theexam: #1}\\ }{\medskip}
\newenvironment{ex}{\noindent{\it Example:} }{\medskip}
\newenvironment{rem}{\noindent{\it Remark:} }{\medskip}

\newcommand{\case}[2]{{\textstyle \frac{#1}{#2}}}
\newcommand{\lP}{\ell_{\mathrm P}}

\newcommand{\md}{{\mathrm{d}}}
\newcommand{\Kern}{\mathop{\mathrm{ker}}}
\newcommand{\tr}{\mathop{\mathrm{tr}}}
\newcommand{\sgn}{\mathop{\mathrm{sgn}}}

\newcommand*{\R}{{\mathbb R}}
\newcommand*{\N}{{\mathbb N}}
\newcommand*{\Z}{{\mathbb Z}}
\newcommand*{\Q}{{\mathbb Q}}
\newcommand*{\C}{{\mathbb C}}

\begin{abstract}
 In many situations, one can approximate the behavior of a quantum
 system, i.e.\ a wave function subject to a partial differential
 equation, by effective classical equations which are ordinary
 differential equations. A general method and geometrical picture is
 developed and shown to agree with effective action results, commonly
 derived through path integration, for perturbations around a harmonic
 oscillator ground state. The same methods are used to describe
 dynamical coherent states, which in turn provide means to compute
 quantum corrections to the symplectic structure of an effective
 system.
\end{abstract}

\section{Introduction}

Many applications of quantum systems are placed in a realm close to
classical behavior, where nevertheless quantum properties need to be
taken into account. In view of the more complicated structure of
quantum systems, both of a conceptual as well as technical nature, it
is then often helpful to work with equations of classical type, i.e.\
systems of ordinary differential equations for mechanical systems,
which are amended by correction terms resulting from quantum
theory. From a mathematical point of view, the question arises how
well the behavior of a (wave) function subject to a partial
differential equation can be approximated by finitely many variables
subject to a system of coupled but ordinary differential equations.

One very powerful method is that of effective actions
\cite{PositronEffAc,VacPolEffAc} which have been
developed and are widely used for quantum field theories. The
effective action of a free field theory is identical to the classical
action, while interacting theories receive quantum corrections ``from
integrating out irrelevant degrees of freedom.'' The language is
suggestive for the physical intuition behind the formalism, but the
technical details and the mathematical relation between classical and
quantum theories remain less clear.

In this article we develop, building on earlier work
\cite{GeomQuantMech,ClassQuantMech,Schilling,Perturb,Josh}, a
geometrical picture of effective equations of motion for a quantum
mechanical system with a clear-cut relation between the classical and
quantum system: as a manifold, a classical phase space of the form
$\R^{2n}$ can literally be embedded into the quantum
system.\footnote{Using the geometrical picture of
\cite{GeomQuantMech,ClassQuantMech} for this purpose and the idea of
horizontality as well as the appearance of additional quantum degrees
of freedom in this context were suggested to us by Abhay Ashtekar.} 
Also the Scchr\"odinger equation can be formulated as Hamiltonian
equations of motion for quantum phase space variables, and
self-adjoint operators as observables in quantum theory select special
functions on the quantum phase space which can be considered as
observables of the classical type. We discuss several examples and
show that, in the regime where effective action techniques can be
used, they coincide with our method.

\section{Effective Actions}

For any system with classical action $S[q]$ as a functional of the
classical coordinates $q$, thus satisfying
\begin{equation}
 \frac{\delta S}{\delta q}=-J
\end{equation}
in the presence of an external source $J$, one can formally define the
effective action $\Gamma[q]$ satisfying the same relation
\begin{equation}\label{EffEq}
 \frac{\delta \Gamma}{\delta q}=-J
\end{equation}
but containing $\hbar$-dependent quantum corrections. If the
generating functional $Z[J]$ of Greens functions is known, $\Gamma$ is
obtained as the Legendre transform of $-i\hbar\log Z[J]$
\cite{EffAcQFT}.

This procedure is well-motivated from particle physics where
additional contributions to $\Gamma$ can be understood as resulting
from perturbative quantum interactions (``exchange of virtual
particles''). Indeed, effective actions are mostly used in
perturbative settings where the generating functional $Z$ can be
computed by perturbing around free theories, using e.g.\ Gaussian path
integrations.

For other systems, or quantum mechanical applications,
Eq.~(\ref{EffEq}) can, however, be seen at best as a formal
justification. The effective action can rarely be derived in general,
but its properties can make an interpretation very complicated. First,
$\Gamma$ is in general complex and so are the effective equations
(\ref{EffEq}) as well as their solutions. In fact, $q$ in
(\ref{EffEq}) is not the classical $q$ and not even the expectation
value of $\hat{q}$ in a suitable state of the quantum system. Instead,
in general it is related to non-diagonal matrix elements of $\hat{q}$
\cite{VariationalEffAc}. Secondly, $\Gamma$ is in general a non-local
functional of $q$ which cannot be written as the time integral of a
function of $q$ and its derivatives. In most applications, one employs
a derivative expansion assuming that higher derivatives of $q$ are
small. In this case, each new derivative order introduces additional
degrees of freedom into the effective action which are not classical,
but whose relation to quantum properties of, e.g., the wave function
is not clear either. Indeed, in this perturbative scheme, not all
solutions of the higher derivative effective action are consistent
perturbatively as many depend non-analytically on the perturbation
parameter $\hbar$ \cite{Simon}. For those solutions, it is then not
guaranteed that they capture the correct perturbative behavior
considering that next order corrections, non-analytical in the
perturbation parameter, can dominate the leading order. Such
non-analytical solutions have to be excluded in a perturbative
treatment, which usually brings down the number of solutions to the
classical value even if perturbative corrections are of higher
derivative form \cite{Simon}. The description, even in a local
approximation, can thus be quite complicated, given by higher
derivative equations with many general solutions subject to the
additional condition that only solutions analytic in the perturbation
parameter are to be retained. The formulation is thus very redundant
if higher derivative terms are used. Moreover, where there seem to be
additional (quantum) degrees of freedom associated with higher
derivative corrections, their role remains dubious given that many
solutions have to be excluded.

There are other technical difficulties if one tries to generalize
beyond the usual realm of perturbing around the ground state of a free
field theory or, in quantum mechanics, the ground state of a harmonic
oscillator. In the latter case, for a system with classical
action
\begin{equation}
 S[q(t)]=\frac{1}{2}m\dot{q}^2-\frac{1}{2}m\omega^2q^2-U(q)
\end{equation}
one can derive the effective action \cite{EffAcQM} (see also
Ref.~\cite{SymmBreakEffAc} for the effective potential)
\begin{equation} \label{EffAc}
\Gamma_{{\rm eff}}[q(t)]=\int
\md t\left[\frac{1}{2}\left(m+\frac{\hbar U'''(q)^2}
{2^5m^2\left(\omega^2+m^{-1}U''(q)\right)^{\frac{5}{2}}}\right)\dot
q^2
-U(q)-\frac{\hbar\omega}{2} \left(1+\frac{U''(q)}{m\omega^2}
\right)^{\frac{1}{2}}\right]
\end{equation}
to first order in $\hbar$ and in the derivative expansion, using path
integral techniques. The quantum system is here described effectively
in an expansion around the ground state of the Harmonic oscillator. On
the other hand, a quantum system allows more freedom and one could,
e.g., want to find an effective formulation for a quantum system which
is prepared to be initially close to a squeezed state, or a state of
non-minimal uncertainty. This freedom is not allowed by the usual
definition of an effective action.

Other problems include the presence of ``infrared problems'': In the
free particle limit, corresponding to a massless field theory, one has
$U(q)=-\frac{1}{2}m\omega^2q^2$ for which (\ref{EffAc}) becomes
meaningless. Still, at least for some time the free particle should be
possible to be described in an effective classical manner. Other
generalizations, such as for systems to be perturbed around a
Hamiltonian non-quadratic in momenta as they occur, e.g., in quantum
cosmology, look even more complicated since one could not rely on
Gaussian path integrations.

For all these reasons it is of interest to develop a scheme for
deriving effective equations of a quantum system based on a
geometrical formulation of quantum mechanics. This has been used
already in the context of quantum cosmology \cite{Perturb,Josh} where
usual techniques fail. As we show here, it reduces to the effective
action result (\ref{EffAc}) in the common range of applicability, but
is much more general. Moreover, it provides a clear, geometrical
picture for the relation between the dynamics of classical and quantum
systems, the role of quantum degrees of freedom and the effective
approximation.

\section{A Geometrical Formulation of Quantum Mechanics}

The formalism of quantum theory has been studied for almost a century
already and a prominent understanding of its structure, based mainly
on functional analysis, has been achieved. From this perspective,
quantum mechanics appears very different from classical mechanics not
only conceptually but also mathematically. While in classical physics
the viewpoint is geometrical, employing symplectic or Poisson
structures on a phase space, quantum theory is analytical and based on
Hilbert space structures and operator algebras.  There are, however,
some contributions which develop and pursue a purely geometrical
picture of quantum mechanics, in which the process of quantization and
kinematical as well as dynamical considerations are generalizations of
classical structures. The process of quantization is described in a
geometrical, though not always constructive, manner in geometric
quantization \cite{Woodhouse}, employing line bundles with
connections, but the picture of the resulting theory remains
analytical based on function spaces and operators
thereon. Independently, a geometrical formulation of quantum mechanics
has been developed which, irrespective of the quantization procedure,
provides a geometrical viewpoint for all the ingredients necessary for
the basic formulation of quantum physics
\cite{GeomQuantMech,ClassQuantMech}. It is the latter which will be
crucial for our purposes of developing a geometrical theory of
effective equations of motion and the classical limit.

Let us assume that we are given a quantum system, specified by a
Hilbert space ${\cal H}=({\cal V},\langle\cdot,\cdot\rangle)$ with
underlying vector space ${\cal V}$ equipped with inner product
$\langle\cdot,\cdot\rangle$, together with an algebra of basic
operators and a Hamiltonian $\hat{H}$. The Hamiltonian defines a flow
on ${\cal H}$ by $\frac{\md\Psi}{\md t}=-i\hbar^{-1}\hat{H}\Psi$.

\begin{lemma}
 Let $({\cal V},\langle\cdot,\cdot\rangle)$ be a Hilbert space. The
inner product $\langle\cdot,\cdot\rangle$ on ${\cal H}$ defines a
K\"ahler structure on ${\cal V}$.
\end{lemma}
\begin{proof}
To start with we note that the inner product can be decomposed as
\begin{equation}
\langle\Phi, \Psi
\rangle=\frac{1}{2\hbar}G(\Phi,\Psi)+\frac{i}{2\hbar}\Omega(\Phi,\Psi)
\end{equation}
where $G(\Phi,\Psi)$ and $\Omega(\Phi,\Psi)$ denote the real and
complex parts of $2\hbar\langle\Phi, \Psi\rangle$, respectively. It
follows from the properties of an inner product that $G$ is a metric
and $\Omega$ a symplectic structure on the vector space
${\cal V}$, identified with its tangent space in any of its
points. Also by definition, the metric and symplectic structure are
related to each other by
\[
 G(\Phi,\Psi)=\Omega(\Phi,i\Psi)\,.
\]
With the obvious complex structure, $({\cal V},G,\Omega)$ is thus
K\"ahler.
\end{proof}

As used in the proof, points and tangent vectors of the K\"ahler
manifold ${\cal K}=({\cal V},G,\Omega)$ correspond to states in the
Hilbert space, and functions densely defined on ${\cal V}$ can be
associated to mean values of operators acting on ${\cal H}$: Any
operator $\hat{F}$ on ${\cal H}$ defines a function $F:=\langle\hat
F\rangle$ on ${\cal K}$ taking values $F(\Psi)=\langle\Psi,\hat
F\Psi\rangle$ in points $\Psi$ of its domain of definition.

Any state $\eta\in{\cal H}$ defines a constant vector field on ${\cal
K}$, which can be used to compute the Lie derivative
\begin{equation}
\pounds_\eta F(\Psi):=\frac{\md}{\md t}F(\Psi+t\eta)|_{t=0}\,.
\end{equation}
This allows us to show
\begin{lemma}
 Let $F=\langle\hat{F}\rangle$ be a function on ${\cal K}$ associated
 with a self-adjoint operator $\hat{F}$ on ${\cal H}$. Its Hamiltonian
 vector field is given by
\[
 X_F(\Psi):=\frac{1}{i\hbar}\hat F\Psi\,.
\]
\end{lemma}
\begin{proof}
Using the definition of a Lie derivative and self-adjointness of
$\hat{F}$ we have
\begin{eqnarray}
\pounds_\eta F(\Psi)&=&\frac{\md}{\md t}\langle\Psi+t\eta,\hat
F(\Psi+t\eta)\rangle|_{t=0}=\langle\eta,\hat
F\Psi\rangle+\langle\Psi,\hat
F\eta\rangle\\\nonumber&=&-i\hbar\left(\langle-i\hbar^{-1}\hat
F\Psi,\eta\rangle-\langle\eta,-i\hbar^{-1}\hat
F\Psi\rangle\right)=\Omega\left(-i\hbar^{-1}\hat
F\Psi,\eta\right)
\end{eqnarray}
for any vector $\eta$, from which $X_F$ can immediately be read off.
\end{proof}

\begin{rem}
Such vector fields are also known as Schr\"odinger vector fields, as
their flow is generated on ${\cal H}$ by a Schr\"odinger equation
\begin{equation}
 \frac{\md}{\md t}|\Psi\rangle=\frac{1}{i\hbar}\hat
F|\Psi\rangle\,.
\end{equation}
The flow is a family of unitary transformations, i.e.\ automorphisms
of the Hilbert space which preserve the Hilbert space
structure. Therefore, the flow preserves not only the symplectic
structure of ${\cal K}$, as any Hamiltonian vector field does, but
also the metric. Hamiltonian vector fields thus are Killing vector
fields, and since each tangent space has a basis of Killing vectors
the K\"ahler space is maximally symmetric.
\end{rem}

For two functions $F=\langle \hat F\rangle$ and $K=\langle\hat
K\rangle$ the symplectic structure defines the Poisson bracket
\begin{equation}
\left\{F,K\right\}:=\Omega(X_F,X_G)=\frac{1}{i\hbar}
\langle[\hat{F},\hat{K}]\rangle\,.
\end{equation}
For, e.g., $q:=\langle\hat q\rangle$ and $p:=\langle \hat p\rangle$ we
have $\left\{q,p\right\}=1$ from $[\hat q,\hat p]=i\hbar$.

Of physical significance in quantum theory are only vectors of the
Hilbert space up to multiplication with a non-zero complex
number. Physical information is then not contained in the vector space
${\cal V}$ but in the projective space ${\cal V}/{\mathbb C}^*$.
From now on we will take this into account by working only with norm
one states and norm-preserving vector fields..

\section{Classical and Quantum Variables}

For any quantum system, the algebra of basic operators, which is a
representation of the classical algebra of basic phase space variables
defined by Poisson brackets, plays an important role. We will assume
mainly, for simplicity, that this basic algebra is given by a set of
position and momentum operators, $\hat{q}^i$ and $\hat{p}_i$ for
$1\leq i\leq N$, with canonical commutation relations. This
distinguished set of operators leads to further structure on ${\cal
K}$:
\begin{defi}
 The set of fundamental operators $(\hat{q}^i,\hat{p}_i)$ on ${\cal
 H}$ defines a fiber bundle structure on ${\cal V}$ where the bundle
 projection identifies all points $\Phi$, $\Psi$ for which
 $\langle\Psi,\hat{q}^i\Psi\rangle=\langle\Phi,\hat{q}^i\Phi\rangle$
 and {$\langle\Psi,\hat p_i\Psi\rangle=\langle\Phi,\hat
 p_i\Phi\rangle$} for all $i$. The base manifold can be identified
 with the classical phase space as a manifold.
\end{defi}

\begin{rem}
 The Hilbert space used for the quantization of a classical system is
 always infinite dimensional, which implies that the fibers of the
 bundle are infinite dimensional.  For instance, for an analytic wave
 function one can consider the collection of numbers associated to the
 mean values of products of the fundamental operators,
 $a_n=\langle\Psi,\hat{q}^n\Psi\rangle$ and $b_n=\langle\Psi,\hat{q}^n
 \hat p\Psi\rangle$ for all $n\geq 0$.  Usually denominated by the
 name of Hamburger momenta \cite{ReedSimon}, the $(a_n,b_n)$ are a
 complete set in the sense that they uniquely determine the wave
 function. Indeed, from linear combinations $c_n$ of the Hamburger
 momenta with coefficients corresponding to some orthogonal
 polynomials, taking Hermite polynomials $\{H_n(q)=\sum_lh_{n,l}q^l\}$
 for definiteness, we have
\begin{equation}
 c_n=\sum_lh_{n,l}a_l=\int dq|\Psi(q)|^2H_n(q)
\end{equation}
giving the absolute value of the wave function as
\begin{equation}
 |\Psi(q)|^2=e^{-q^2}\sum_n \frac{c_nH_n(q)}{2^n\pi n!}\,.
\end{equation}
The $b_n$, on the other hand, provide information about
the phase $\alpha(q)$ of the wave function up to a constant:
\begin{equation}
 b_n=-\int dq\Psi(q)^*{q}^n
 i\partial_q\Psi(q)=-\int \md q|\Psi(q)|{q}^n i\partial_q|\Psi(q)|-\int
 \md q|\Psi(q)|^2{q}^n i\partial_q\alpha(q)
\end{equation}
from which $\partial_q\alpha(q)$ is determined as before, using the
already known norm of $\Psi$.
\end{rem}

One could thus use Hamburger momenta as coordinates on the fiber
bundle, but for practical purposes it is more helpful to choose
coordinates which are not only adapted to the bundle structure but
also to the symplectic structure. We thus require that, in addition to
the classical variables $q^i$ and $p_i$, coordinates of the fibers
generate Hamiltonian vector fields symplectically orthogonal to
$\partial/\partial q^i$ and $\partial/\partial p_i$.

\begin{defi}
The {\em quantum variables} of a Hilbert space ${\cal H}$ are defined as
\begin{equation}
 G^{i_1 \ldots i_n}\!:=\! \langle(\hat{x}^{(i_1}\!-x^{(i_1}) \cdots
 (\hat{x}^{i_n)}\!-x^{i_n)})\rangle
 =\sum_{k=0}^{n}(-)^k(_k^n)x^{(i_1}\!\! \cdots x^{i_k}
 \langle\hat{x}^{i_{k+1}}\!\! \cdots \hat{x}^{i_n)}\rangle
\end{equation}
with respect to fundamental operators
$\{\hat{x}^i\}_{1\leq i\leq 2N}:=\{\hat{q}^k,\hat{p}_k\}_{1\leq i\leq N}$
where round brackets on indices denote symmetrization.
\end{defi}

Variables of this type have been considered in quantum field theories;
see, e.g., \cite{EffAcComp}. Together with the classical variables,
they provide in particular local trivializations of the quantum phase
space as a fiber bundle.

\begin{lemma}
 The fiber coordinates $G^{i_1 \ldots i_n}$ on ${\cal K}$ are
symplectically orthogonal to the classical coordinates
$x^i$.
\end{lemma}
\begin{proof}
 We compute the Poisson bracket with $x^j$ to obtain
\begin{eqnarray}
 \left\{x^j,G^{i_1 \cdots i_n}\right\} &=&
 \sum^{n}_{k=0}(-)^k(^n_k)\left[\left\{x^j,x^{(i_1} \cdots
 x^{i_k}\right\} \langle\hat x^{i_{k+1}} \cdots \hat x^{i_n)}\rangle
+x^{(i_1} \cdots x^{i_k}\left\{x^j, \langle\hat x^{i_{k+1}} \cdots
\hat x^{i_n)}\rangle\right\}\right]\nonumber\\
&=&\sum_{k=0}^n
(-)^k(^n_k)\left[k\epsilon^{j(i_n}x^{i_1}\!\! \cdots x^{i_{k-1}}
\langle\hat{x}^{i_{k}}\!\! \cdots \hat{x}^{i_{n-1})}\rangle+
(n-k)\epsilon^{j(i_n}x^{i_1}\!\! \cdots x^{i_k}
\langle\hat{x}^{i_{k+1}}\!\! \cdots
\hat{x}^{i_{n-1})}\rangle\right]\nonumber\\
&=&\sum_{l=0}^{n-1}
(-)^{(l+1)}(^n_l) (n-l)\epsilon^{j(i_n}x^{i_1}\!\! \cdots x^{i_l}
\langle\hat{x}^{i_{l+1}}\!\! \cdots
\hat{x}^{i_{n-1})}\rangle\nonumber\\
&&+\sum_{k=0}^{n} (-)^k(^n_k)
(n-k)\epsilon^{j(i_n}x^{i_1}\!\! \cdots x^{i_k}
\langle\hat{x}^{i_{k+1}}\!\! \cdots \hat{x}^{i_{n-1})}\rangle=0
\end{eqnarray}
where we used repeatedly the
Leibnitz rule and introduced
$\epsilon^{ij}=\left\{x^i,x^j\right\}$.
\end{proof}

\begin{rem}
 An alternative proof proceeds by computing the Poisson bracket
 between the function $\langle e^{\alpha_i(\hat x^i-x^i)}\rangle$ and
 $x^j$, restricting to the dense subspace in which such functions are
 analytic in $\{\alpha_i\}$, and expanding.
\end{rem}

Since the fibers are symplectic, $\Omega$ defines a natural
decomposition of tangent spaces of ${\cal K}$ as a direct sum of a
vertical space tangent to the fibers and a horizontal space ${\rm
Hor}^{\Omega}{\cal K}$ as the symplectic complement:
\begin{corr}
 $({\cal K},\pi,{\cal B})$ is a fiber bundle with connection over the
classical phase space ${\cal B}$ as base manifold.
\end{corr}

We now know the Poisson relation between the classical variables $x^i$
and between $x^i$ and the $G^{j_1,\ldots, j_m}$. In order to compute
the remaining Poisson brackets $\left\{G^{i_1, \ldots,
i_n},G^{j_1,\ldots, j_m}\right\}$ for $N$ canonical degrees of freedom
we introduce a new notation
\begin{equation}
G^{a_{k_1},\dots ,a_{k_N}}_{b_{k_1},\dots, b_{k_N}}= \langle(\hat
q^{k_1}-q^{k_1})^{a_{k_1}}\cdots(\hat q^{k_N}-q^{k_N})^{a_{k_N}}(\hat
p_{k_1}-p_{k_1})^{b_{k_1}}\cdots(\hat
p_{k_N}-p_{k_N})^{b_{k_N}}\rangle_{\rm Weyl}\nonumber
\end{equation}
the label ``Weyl'' meaning that the product of operators is Weyl or
fully symmetric ordered. The notation allows us to drop indices whose
values are zero so whenever we are dealing with a single pair of
degrees of freedom we use the notation where $G^{a,n}:=G^{n-a}_a$.
\begin{lemma}
The Poisson brackets for the variables above are
\begin{eqnarray}\nonumber\left\{G^{a_{k_1},\dots
,a_{k_N}}_{b_{k_1},\dots,b_{k_N}},G^{c_{k_1},\dots,c_{k_N}}_{d_{k_1},
\dots,d_{k_N}}\right\}\!\!
=\!\!-\!\!\!\!\!\!\!\!\sum_{r,s,{e_1,\dots,e_N}}\!\!\!\!\!\!
(-)^{r+s}({\textstyle\frac{1}{2}}\hbar)^{2r}
\delta_{e_1+\dots+e_N,2r+1}K_{r,s,\{e\}}^{\{a\}\{b\}\{c\}\{d\}}
G^{a_{k_1}+c_{k_1}-e_1,
\dots,a_{k_N}+c_{k_N}-e_N}_{b_{k_1}+d_{k_1}-e_1,\dots,b_{k_N}+
d_{k_N}-e_N}&&\\\nonumber-\sum_{f=1}^N\!\!
\left( a_{k_{\!f}}d_{k_{\!f}}G^{a_{k_1},\dots, a_{k_{\!f}}-1,\dots,
a_{k_N}}_{b_{k_1},\dots,b_{k_N}}G^{c_{k_1},\dots,c_{k_N}}_{d_{k_1},\dots,
d_{k_{\!f}}-1,\dots, d_{k_N}}\!\!\!-\!
b_{k_{\!f}}c_{k_{\!f}}G^{a_{k_1},\dots,a_{k_N}}_{b_{k_1},\dots,b_{k_{\!f}}-1
,\dots,b_{k_N}}G^{c_{k_1},\dots,c_{k_{\!f}}-1,\dots,c_{k_N}}_{d_{k_1},\dots,
d_{k_N}}\right)&&
\end{eqnarray}with indices running as
\begin{eqnarray}&1\leq
2r+1\leq\sum_{f=1}^N\left(\min(a_{\!f},d_{\!f})+
\min(b_{\!f},c_{\!f})\right),&\nonumber
\\&0\leq s\leq\min\left(r,\sum_{f=1}^N\min(b_{\!f},c_{\!f})\right),&\nonumber\\
&0\leq
e_{\!f}\leq\min(a_{\!f},d_{\!f},s)+\min(b_{\!f},c_{\!f},2r+1-s).&
\nonumber\end{eqnarray} and coefficients given by
\begin{equation}
 K_{r,s,\{e\}}^{\{a\}\{b\}\{c\}\{d\}}=
\sum_{g_1,\dots,g_n}
\frac{\delta_{g_1+\dots+g_N,2r+1-s}}{s!(2r+1-s)!}\prod_f\frac{
\left(^{\ \ a_{\!f}}_{e_{\!f}-g_{\!f}}\right)
\left(^{b_{\!f}}_{g_{\!f}}\right)\left(^{c_{\!f}}_{g_{\!f}}\right)\left(^{\
\ d_{\!f}}_{e_{\!f}-g_{\!f}}\right)}{\left(^{ 2r+1-s}_{\ \
g_{\!f}}\right)\left(^{\ \ s}_{e_{\!f}-g_{\!f}}\right)}
\end{equation}
where
\[
\max(e_{\!f}-s,e_{\!f}-a_{\!f},e_{\!f}-d_{\!f},0)\leq g_{\!f}\leq
\min(b_{\!f},c_{\!f},2r+1-s,e_{\!f})\,.
\]
\end{lemma}

\begin{proof}
 Consider first the Poisson bracket between functions of the form
 $D(\alpha)=\langle e^{\alpha_i(\hat x^i -x^i)}\rangle$.  For
 analytical wave functions in the mean values, $D(\alpha)$ is an
 analytical function and so is the Poisson bracket between two such
 functions $D(\alpha)$ and $D(\beta)$.  We can therefore take the
 coefficients in a Taylor expansion for all orders in $\alpha_i$ and
 $\beta_j$. Using the relation $[e^{\alpha_i\hat x^i}, e^{\beta_j\hat
 x^j}]=2 i
 \sin(\frac{\hbar}{2}\alpha_j\beta_k\epsilon^{jk})e^{(\alpha+\beta)_i\hat
 x^i}$, which follows from the Baker--Campbell--Hausdorff formula and
 the commutator $[\alpha_i\hat{x}^i,\beta_j\hat{x}^j]=
 i\hbar\epsilon^{ij} \alpha_i\beta_j$, we find that
\begin{equation}\label{symplecticstructureD}
 \left\{D(\alpha),D(\beta)\right\}=\frac{2}{\hbar}
 \sin({\textstyle\frac{1}{2}}\hbar\alpha_j\beta_k\epsilon^{jk})D(\alpha+\beta)
 -\alpha_j\beta_k\epsilon^{jk}D(\alpha)D(\beta)\,.
\end{equation}
Now, we use $D(\alpha)=\langle e^{\alpha_i(\hat x^i -x^i)}\rangle=
\underset{\{a\},\{b\}}{\sum}G^{a_1\cdots a_N}_{b_1\cdots b_N}
\prod_{i=1}^N\alpha_{q^i}^{a_i}\alpha_{p_i}^{b_i}(a_i!b_i!)^{-1}$,
and substitute
\begin{eqnarray}
 \frac{2}{\hbar}\sin({\textstyle\frac{1}{2}}\hbar\alpha_j\beta_k\epsilon^{jk})
 D(\alpha+\beta)&=&
 -\sum(-)^{r+s}({\textstyle\frac{1}{2}\hbar})^{2r}G^{a_1+c_1,\cdots
 ,a_N+c_N}_{b_1+d_1,\cdots,b_N+d_N}\\
 &&\times\prod_{f=1}^N\frac{\alpha_{q^f}^{a_f+g_f}\alpha_{p_f}^{b_f+e_f}
\beta_{q^f}^{c_f+e_f}\beta_{p_f}^{d_f+g_f}
 }{a_f!b_f!c_f!d_f!e_f!g_f!(2r+1-s-e_f)!(s-g_f)!}
\nonumber
\end{eqnarray}
where we sum over all collections of numbers $a_f$, $b_f$, $c_f$,
$d_f$, $e_f$, $g_f$, $r$ and $s$ such that $\sum_f g_f=s$, $\sum_f
e_f=2r-s$ and $s\leq 2r+1$. Since the equality
(\ref{symplecticstructureD}) holds for any $\alpha$ and $\beta$,
coefficients  in the expansion have to fulfill the equality.
\end{proof}

\section{Uncertainty Principle}

The fibers of ${\cal K}$ as a fiber bundle over the classical phase
space are not vector spaces, and the quantum variables $G^{i_1,\ldots,
i_n}$ are not allowed to take arbitrary values. Similarly, not any
collection of numbers is a collection of Hamburger momenta. With
${\cal K}$ being a K\"ahler space, the fibers are bounded by relations
following from Schwarz inequalities.  A special case of this fact is
well-known and commonly written as the uncertainty relation
\begin{equation}
 (\triangle
 q )^2(\triangle p)^2\geq\frac{\hbar^2}{4}+\langle(\hat q\hat
 p+\hat p\hat q)/2-qp\rangle^2\geq \frac{\hbar^2}{4}
\end{equation}
where $(\triangle a)^2=\langle(\hat a -a)^2\rangle$, or in our
notation
\begin{equation} \label{UncertG2}
 G^{0,2}G^{2,2}\geq\frac{\hbar^2}{4}+(G^{1,2})^2\,.
\end{equation}

More generally, the Schwarz inequality for a K\"ahler manifold with
metric $g$ and symplectic structure $\omega$ is
\begin{equation}
 g(u,u)g(v,v)\geq
\left|g(u,v)\right|^2+\left|\omega(u,v)\right|^2
\end{equation}
for all tangent vectors $u$ and $v$.
This results in bounds to be imposed on the quantum variables.
\begin{lemma}
 The function $D(\alpha)=\langle e^{\alpha_i(\hat{x}^i-x^i)}\rangle$
 is subject to
\begin{equation}\label{uncertainty}
 \left(\!D(2\alpha)\!\!-\!\!D(\!\alpha)^2\right)\!\left(\!D(2\beta)\!-\!
 D(\beta)^2\!\right)\geq
 D(\!\alpha+\!\beta)^2\!-\!2\cos({\textstyle\frac{1}{2}\hbar}
 \alpha\times\beta)
 D(\!\alpha+\beta)D(\!\alpha)D(\beta)+D(\!\alpha)^2D(\beta)^2\,.
\end{equation}
\end{lemma}
\begin{proof}
 For the Schwarz inequality we need to know the metric and
 pre-symplectic structure on the space of states of unit norm, which
 we compute by evaluating them on vector fields that generate
 transformations only along the submanifold of unit vectors in the
 Hilbert space. To an arbitrary vector $X_F=\frac{1}{i\hbar}\hat
 F\Psi$ we associate the vector given by $\tilde X_F=\left(1-
 |\Psi\rangle \langle\Psi|\right)X_F = \frac{1}{i\hbar}(\hat F-
 F)\Psi$.
This ensures that the transformation
 generated by $\tilde X_F$ maps normalized states to normalized
 states, which is most easily seen infinitesimally using
 $|(1+\epsilon\tilde{X}_F)\Psi|^2=|\Psi|^2-2i\hbar^{-1}\epsilon
 \langle\Psi,(\hat{F}-F)\Psi\rangle+O(\epsilon^2)=|\Psi|^2+O(\epsilon^2)$.
 The metric on the space of physical states evaluated in Hamiltonian
 vector fields induces a symmetric bracket
\begin{equation}
 (F,K)=g(X_F,X_K)=G(\left(1-
 |\Psi\rangle \langle\Psi|\right)X_F,\left(1- |\Psi\rangle
 \langle\Psi|\right)X_K)\,.
\end{equation}
The symplectic structure is as before,
$\omega(X_F,X_K)=\Omega(X_F,X_K)$.
For the corresponding operators, $g$ and $\omega$
result in the anticommutator $[\cdot,\cdot]_+$ and commutator
$[\cdot,\cdot]$, respectively.

For functions $\langle e^{\alpha.\hat x}\rangle$ and $\langle
e^{\beta.\hat x}\rangle$ (parameterized by $\alpha_i$ and $\beta_i$)
the Schwarz inequality implies
\begin{equation}
 (\langle e^{2\alpha.\hat x}\rangle-\langle
e^{\alpha.\hat x}\rangle^2)(\langle e^{2\beta.\hat x}\rangle-\langle
e^{\beta.\hat
x}\rangle^2)\geq\left|\frac{1}{2}
\langle[e^{\alpha.\hat x},e^{\beta.\hat x}]_+\rangle-\langle
e^{\alpha.\hat x}\rangle\langle e^{\beta.\hat x}\rangle\right|^2
+\frac{1}{4}\left|\langle[e^{\alpha.\hat
x},e^{\beta.\hat x}]\rangle\right|^2
\end{equation}
which upon using, as before, the Baker--Campbell--Hausdorff formula
for the commutator and anticommutator and multiplying both sides with
$e^{-2(\alpha+\beta).x}$ proves the lemma.
\end{proof}

This gives us a large class of inequalities thus specifying bounds
on the variables $G^{i_1,\ldots, i_n}$. The boundary, obtained through
saturation of the inequalities, is characterized by relations which
result from the lemma order by order in $\alpha$ and $\beta$.

\section{Quantum Evolution}

The  dynamical flow of the quantum system is given as the unitary
Schr\"odinger flow on ${\cal H}$ of a self-adjoint Hamiltonian
operator $\hat{H}$. As before, this flow is also Hamiltonian when
viewed on the K\"ahler space ${\cal K}$. It is generated by the
Hamiltonian function obtained as the mean value of the Hamiltonian
operator.  In terms of coordinates on the manifold the Hamiltonian
function is obtained by Taylor expanding the mean value of the
Hamiltonian operator which in our convention is taken to be Weyl
ordered:
\begin{defi}
 The {\em quantum Hamiltonian}\footnote{This is the basic object for
 an effective theory, playing a similar role in the effective
 potential \cite{EffPotExp}.} on ${\cal K}$ is the function
\begin{equation} \label{HQ}
H_Q:=\langle H(\hat{x}^i)\rangle_{\rm Weyl}=\langle
H(x^i+(\hat{x}^i-x^i))\rangle =\sum_{n=0}^{\infty} \sum_{a=0}^n
\frac{1}{n!}\binom{n}{a}\frac{\partial^n
H(q,p)}{\partial p^a\partial q^{n-a}}G^{a,n}
\end{equation}
generating {\em Hamiltonian equations of motion}
\begin{eqnarray}
 \dot{x}^i=&\{x^i,H_Q\}\nonumber\\
 \dot{G}^{a,n}=&\{G^{a,n},H_Q\}\,.
\end{eqnarray}
\end{defi}

This Hamiltonian flow is equivalent to the Schr\"odinger equation of
the Hamiltonian operator. As such, it is an equivalent description of
the quantum dynamics and only superficially takes a classical form,
albeit for infinitely many variables, in its mathematical
structure. Nevertheless, the reformulation makes it possible to
analyze the classical limit in a direct manner, and to derive
effective equations in appropriate regimes.  Classical dynamics is to
arise in the limit of ``small" quantum fluctuations which, when the
fluctuations are completely ignored or switched off by $\hbar\to0$,
should give rise to classical equations of motion.  In practice, this
limit is not easy to define, and the most direct way is to derive
first effective equations of motion, which still contain $\hbar$, and
then take the limit $\hbar\to0$.

In this procedure, the main problem is to reduce the infinite set of
coupled quantum equations of motion to a set of differential
equations for only a finite set of variables.  Additional degrees of
freedom without classical analogs carry information about, e.g., the
spreading of the wave function around the peak, which itself is
captured by expectation values. For a formulation of classical type,
taking into account only a finite number of degrees of freedom, a
system has to allow a {\em finite-dimensional} \,submanifold of the
quantum space ${\cal K}$ which is preserved by the quantum flow. We
start by generalizing the situation encountered in \cite{Schilling}:
\begin{defi} \label{StrongEff}
 A {\em strong effective classical system} $({\cal P},H_{\rm eff})$
 for a quantum system $({\cal H},\hat{H})$ is given by a finite
 dimensional pre-symplectic subspace ${\cal P}$ of the K\"ahler space
 ${\cal K}$ associated with ${\cal H}$ satisfying the following two
 conditions:
\begin{enumerate}
\item For each $p\in{\cal P}\subset{\cal K}$ the tangent space
 $T_p{\cal P}$ contains the horizontal subspace ${\rm
 Hor}_p^{\Omega}{\cal K}$ of $p$ in ${\cal K}$ defined by the
 symplectic structure: ${\rm Hor}_p^{\Omega}{\cal K}\subset T_p{\cal
 P}$ for all $p\in {\cal P}$ {\em (base horizontality)}.
\item ${\cal P}$ is fixed under the
 Schr\"odinger flow of $\hat{H}$ and, if ${\cal P}$ is symplectic, the
 restriction of the flow to ${\cal P}$ agrees with the Hamiltonian
 flow generated by the {\em effective Hamiltonian} $H_{\rm eff}$.
\end{enumerate}
\end{defi}

\begin{rem}
 A strong effective classical system agrees with the quantum system
 both at the kinematical and quantum level since its symplectic
 structure as well as the Hamiltonian flow are induced by the
 embedding. As such, the conditions are very strong since they require
 a quantum system to be described {\em exactly} in terms of a {\em finite
 dimensional} system ${\cal P}$. In addition to agreement between the
 strong effective and the quantum dynamics, the first condition
 ensures that the classical variables are contained in ${\cal P}$ and
 fulfill the classical Poisson relations.
\end{rem}

In the simplest case we require the effective system to have the same
dimension as the classical system, such that potentially only
correction terms will appear in $H_{\rm eff}$ (to be discussed further
in Theorem~\ref{StrongEffHam} below) but no additional degrees of
freedom.  Quantum variables, in general, cannot simply be ignored
since they evolve and back react on the classical variables. Sometimes
one may be forced to keep an odd number of quantum variables, such as
the three $G^{a,2}$, in the system which we allow by requiring the
effective phase space ${\cal P}$ to be only pre-symplectic.  For a
strong effective system of the classical dimension, however, the
dynamics of the quantum variables in the embedding space occurs only
as a functional dependence through the classical coordinates:
\begin{equation}
 \dot{G}^{a,n}=\dot{x}^i\partial_{x^i}G^{a,n}(x^j)\,.
\end{equation}
The effective equations of motion, generated by $H_{\rm eff}$ are then
obtained by inserting solutions
$G^{a,n}(x)$ in the equations for $x^i$:
\begin{equation}
 \dot{x}^i=\{x^i,H_Q\}|_{G^{a,n}(x)}=\sum_{n=0}^{\infty}
\frac{1}{n!}\{x^i,H(x^i),_{i_1\dots i_n}\}G^{i_1,\ldots, i_n}(x)\,.
\end{equation}

\section{Examples}

We now demonstrate the applicability of the general procedure by
presenting examples, which will then lead the way to a weakened
definition and, in the following section, a proof that the results
coincide with standard effective action techniques when both can be
applied.

\begin{excount}{Harmonic Oscillator}
The quantum Hamiltonian (\ref{HQ}) for a harmonic
oscillator is
\begin{equation}
H_Q=\frac{1}{2m}p^2+\frac{1}{2}m\omega^2q^2
+\frac{1}{2}m\omega^2G^{0,2}+\frac{1}{2m}G^{2,2}
\end{equation}
giving equations of motion
\begin{eqnarray}
\nonumber \dot{p}=&\{p,H_Q\}&=-m\omega^2 q\\
\dot{q}=&\{q,H_Q\}&=\frac{1}{m} p\\\nonumber
\dot{G}^{a,n}=&\{{G}^{a,n},H_Q\}&=\frac{1}{m}(n-a)G^{a+1,n}-m\omega^2
aG^{a-1,n}\,.
\end{eqnarray}
In this case, the set of infinitely many coupled equations splits into
an infinite number of sets, for each $n$ as well as the classical
variables, each having a finite number of coupled
equations. Independently of the solutions for the $G^{a,n}$ we obtain
the same set of effective equations for $q$ and $p$ agreeing with the
classical ones. Therefore the effective Hamiltonian for a system of
the classical dimension is here identical to the classical one (up to a
constant which can be added freely).  We can also define higher
dimensional (but non-symplectic) systems by including the variables
$G^{a,n}$ for a finite set of values for $n$.

Along the classical evolution, the evolution of the additional
parameters is then given by linear differential equations which we
write down in a dimensionless form, defining 
\begin{equation} \label{Gtilde}
 \tilde G^{a,n}=\hbar^{-n/2}(m\omega)^{n/2-a}G^{a,n}\,.
\end{equation}
The requirement that
dynamics be restricted to the classical subspace parameterized by
$q$ and $p$ implies
\begin{equation} \label{M}
\frac{1}{\omega}\left(\frac{1}{m}p\partial_q-
m\omega^2q\partial_p\right) \tilde{G}^{a,n}=(n-a)\tilde{G}^{a+1,n}-
a\tilde{G}^{a-1,n}=:{}^{(n)}\!M^a_{\ b}\tilde{G}^{b,n}
\end{equation}
whose solution is
\begin{equation}
 \tilde{G}^{a,n}(r,\theta)=(\exp\theta{}^{(n)}\!M)^a_{\ b}A^b(r)
\end{equation}
where $r=\sqrt{\frac{1}{m}p^2+m\omega^2 q^2}$, $\tan
(\theta)=m\omega q/p$ and $A^{a,n}(r)$ are $n+1$ arbitrary functions
of $r$. For, e.g., $n=2$ we have
\begin{eqnarray}
\tilde{G}^{0,2}(r,\theta)&=&A^{0,2}(r)-e^{2i\theta}A^{2, 2}(r)-
e^{-2i\theta}A^{-2, 2}(r) \label{HOG02}\\
\tilde{G}^{1,2}(r,\theta)&=&-ie^{2i\theta}A^{2,2}(r)+ie^{-2i\theta}
A^{-2,2}(r)\label{HOG12}\\
\tilde{G}^{2,2}(r,\theta)&=&A^{0,2}(r)+e^{2i\theta}A^{2,2}(r)+
e^{-2i\theta}A^{-2,2}(r)\label{HOG22}
\end{eqnarray}
In terms of the constants $A^{a,n}$, the uncertainty relation
(\ref{UncertG2}) reads:
\begin{equation}
 (A^{0,2}(r))^2-4A^{2,2}(r)A^{-2,2}(r)\geq\frac{1}{4}
\end{equation}
We are thus allowed to choose $A^{2,2}=0=A^{-2,2}$ and
$A^{0,2}=\frac{1}{2}$ which saturates the uncertainty bound and
makes the $G^{a,2}$ constant. In fact, these values arise from
quantum evolution given by coherent states $|\alpha\rangle
=e^{\alpha\hat a^\dagger-\bar{\alpha}\hat a}|0\rangle$ which
corresponds to trajectories of {\em constant} quantum variables
\begin{equation}\label{freemoments}
\tilde{G}^{a,n}=\frac{1}{2^n}\frac{a!}{(a/2)!}\frac{(n-a)!}{((n-a)/2)!}
\end{equation}
for even $a$ and $n$, and $\tilde{G}^{a,n}=0$ otherwise. This implies
that any truncation of the system by including only a finite set of
values for $n$, which as already seen is consistent with the dynamical
equations, and choosing initial conditions to be that of a
coherent state gives a base horizontal subspace as required by
Def.~\ref{StrongEff}. In other words, the harmonic oscillator allows
an infinite set of strong effective classical systems, including one
of the classical dimension. The last case is symplectic, with
effective Hamiltonian $H_{\rm eff}=H+{\rm const}$.

In particular, for $n=2$ we see that the uncertainty relations are
saturated. For other states, the quantum variables will in general
vary during evolution, which means that the spreading of states
changes in time. Nevertheless, the variables remain bounded and the
system will stay in a semiclassical regime of small uncertainties if
it starts there. With varying $G$, we will not obtain a strong
effective system as horizontality will be violated. Nevertheless, such
states are often of interest and suitable for an effective
description, which we will provide in a weakened form later on.
\end{excount}

\begin{excount}{Linear systems}
The harmonic oscillator is a special case of systems, where a complete
set of functions on the classical phase space exists such that they
form a Lie algebra with the Hamiltonian. For such systems, which we
call linear, semiclassical aspects can be analyzed in an elegant
manner using {\em generalized coherent states}: a family of states ---
of the dimension of the algebra minus the dimension of its subalgebra
that generates the stability subgroup of a given, so-called extremal
state --- with respect to which the mean values of operators can be
approximated very well by their classical expressions \cite{CohStates}.

In this example we assume that basic variables of the quantum system
are not necessarily canonical but given by the Lie algebra elements
$\hat{L}^i$ of a linear quantum system. Thus, our classical
variables are mean values $L^i:=\langle\hat L^i\rangle$, and quantum
variables are
\[
G_L^{i_1,\ldots, i_n}=\langle(\hat L^{(i_1}-L^{i_1})\cdots(\hat
L^{i_n}-L^{i_n)})\rangle\,.
\]
Poisson brackets between these
functions on the infinite dimensional K\"ahler manifold ${\cal K}$
can easily be found to be
\[
 \{L^i,L^j\}=f^{ij}{}_{k}L^k
\]
and
\[
 \{L^i,G_L^{i_1,\ldots, i_n}\}=\sum_{r,j} f^{ii_r}{}_{j}
 G_L^{i_1,\ldots, i_{r-1}ji_{r+1},\ldots, i_n}\,.
\]

It is then immediately seen that the Hamiltonian dynamics of all
degrees of freedom is linear, the $L^i$ decouple from the quantum
variables, and that the dynamics of any $G_L^{i_1,\ldots, i_n}$
depends only on other $G_L^{j_1,\ldots, j_n}$ with the same $n$. As in
the harmonic oscillator case, the dynamics of infinitely many
degrees of freedom thus decouples into infinitely many sectors
containing only finitely many variables. This shows

\begin{corr} \label{IntegrableCorr}
 Any linear quantum system admits a class of finite dimensional
 subspaces preserved by the quantum flow, including one of the
 classical dimension.
\end{corr}

This is not sufficient for the existence of a strong effective system,
for which we also have to discuss base horizontality. As in the
harmonic oscillator example, one can try to use coherent states which
have been widely analyzed in this context. Nevertheless, the issue of
base horizontality, i.e.\ finding coherent states for which all $G$
are constant, in general is more complicated.

A special family of states is generated by acting with the Lie
algebra on an extremal state, i.e.\ a lowest weight of a module
representation, which can thus be seen to be in one-to-one
correspondence with the factor space of the Lie algebra by the
stabilizer of the state. More explicitly those states are of the
form
\[
 |\eta\rangle_{\Lambda,\Omega} =e^{\sum_\alpha\eta_\alpha E_\alpha-H.c.}|{\rm
ext}\rangle
=N(\tau(\eta),\tau(\eta)^*)^{-1}e^{\sum_\alpha\tau_\alpha(\eta)
E_\alpha}|{\rm ext}\rangle\,,
\]
where $\Lambda$ is a representation of the Lie algebra, $\Omega$ is
the quotient of the Group manifold by its stabilizer, $|{\rm
ext}\rangle$ is an extremal state, $E_{-\alpha}|{\rm ext}\rangle=0$
for all positive roots $\alpha$ and $\eta_\alpha$ or $\tau_\alpha$
are coordinate charts of the homogeneous space. Since the flow is
generated by an element of the Lie algebra, generalized coherent
states define a preserved manifold according to the
Baker--Campbell--Hausdorff formula.

In this situation one can compute the mean values of elements $L^i$ of
the Lie algebra and the quantum variables $G_L^{i_1,\ldots, i_n}$ as
functions over the classical phase space. With this construction of
coherent states, the semiclassical phase space associated to the Lie
algebra and the dimension of the classical theories would differ
depending on the choice of the extremal state and each of these would
provide us with diffeomorphisms from the set of $L^i$ to the
$\tau_\alpha$, these last ones being the only dynamical variables of
this subspace (when all conditions are satisfied, we have by definition
{\em dynamical coherent states}).

We can notice as well that a natural emergence of a K\"ahler structure
for this submanifold of the space of states, as observed within the
context of the geometrical formulation of quantum mechanics, is also
justified in Gilmore's construction.

We are not aware of general expressions for the $G$ or special choices
of constant values as they exist for the harmonic oscillator. It is,
however, clear that such constant choices are not possible in general
for a linear system as the counter-example of the free particle
demonstrates.
\end{excount}

\begin{excount}{Free Particle}
 The free particle is an example for a linear system and can be
 obtained as the limit of a harmonic oscillator for
 $\omega\to0$. However, the limit is non-trivial and the semiclassical
 behavior changes significantly. If we re-instate units into the
 uncertainty formulas of the harmonic oscillator, we obtain in the
 case of constant $G^{a,2}$:
\[
 G^{0,2}=\frac{\hbar}{2m\omega}\quad,\quad G^{1,2}=0\quad,\quad
G^{2,2}=\frac{\hbar{m\omega}}{2}\,.
\]
 The fixed point of the evolution of quantum variables which exists
 for the harmonic oscillator thus moves out to infinity in the free
 particle limit and disappears. Moreover, the closed classical orbits
 break open and become unbounded. Even non-constant bounded solutions
 for the $G$ then cease to exist, a fact well-known from quantum
 mechanics where the wave function of a free particle has a strictly
 growing spread, while harmonic oscillator states always have bounded
 spread as follows from (\ref{HOG02}), (\ref{HOG12}) and
 (\ref{HOG22}). For a free particle, one can thus not expect to have a
 valid semiclassical approximation for all times.

 One can see this explicitly by computing eigenvalues of the matrices
 ${}^{(n)}\!M$ in (\ref{M}) for arbitrary $n$ which in the limit of
 vanishing frequency become degenerate. More precisely, the solutions of
\begin{equation}
 \frac{p}{m}\partial_qG^{a,n}=\frac{n-a}{m}G^{a+1,n}
\end{equation}
are given by
\begin{equation}
 G^{a,n}(q,p)=p^a\sum_{i=0}^{n-a}\frac{c_{i,n}(n-a)!}{(n-a-i)!}q^{n-a-i}
\end{equation}
 with integration constants $c_{i,n}$, $i=0,\ldots,n$.  Minimal
 uncertainty requires for $n=2$ that
 $2c_0c_2-c_1^2=\frac{\hbar^2}{4p^2}$. Initial conditions could be
 chosen by requiring the initial state to be a Harmonic oscillator
 coherent state at the point $(q_0,p_0)$. Since, due to the degeneracy
 of eigenvalues, solutions for the $G$ are now polynomials in $q$ and
 the classical trajectories are unbounded, the spread is unbounded
 when the whole evolution is considered. In particular, no constant
 choice and so no strong effective system exists. With unbounded
 quantum variables, the system cannot be considered semiclassical for
 all times, but for limited amounts of time this can be reasonable. If
 this is done, the equations of motion for the classical variables $q$
 and $p$ are unmodified such that there is no need for introducing an
 effective Hamiltonian different from the classical one if one is
 interested only in an effective system of the classical dimension.

\end{excount}

\begin{excount}{Quantum Cosmology}
 So far we have mainly reproduced known results in a different
 language. To illustrate the generality of the procedure we now
 compute effective equations for an unbounded Hamiltonian which
 generally occurs in quantum cosmology. Here, one considers the
 quantized metric of a homogeneous and isotropic space-time whose sole
 dynamical parameter is the scale factor $a$ determining the change of
 size of space in time. The canonical structure as well as Hamiltonian
 follow from the Einstein--Hilbert action specialized to such an
 isotropic metric. The momentum is then given by
 $p_a=3a\dot{a}/\kappa$ with the gravitational constant $\kappa$, and
 the Hamiltonian is equivalent to the Friedmann equation. There are
 different sets of canonical variables, all related to the spatial
 metric and extrinsic curvature of spatial slices, some of which are
 better adapted to quantization. Here, we use the example of isotropic
 quantum cosmology coupled to matter in the form of dust (constant
 matter energy $E$) in Ashtekar variables \cite{AshVar} which in the
 isotropic case are $(c,p)$ with $\{c,p\}=\frac{1}{3}\gamma\kappa$
 where $\gamma$ is a real constant, the so-called Barbero--Immirzi
 parameter\cite{AshVarReell,Immirzi}, and give a Hamiltonian
 $H=-3\gamma^{-2}\kappa^{-1}c^2\sqrt{p}+E$. (This is formally similar
 to a system with varying mass as discussed in \cite{EffAcVarMass}.)
 For details of the variables $(c,p)$ used we refer to
 \cite{IsoCosmo,Bohr}. The geometrical meaning can be seen from
 $|p|=a^2$ and $c=\frac{1}{2}\gamma\dot{a}$ in terms of the scale
 factor $a$. For a semiclassical universe, we thus have $c\ll1$ and
 $p\gg \ell_{\rm P}^2=\hbar\kappa$ In contrast to $a$, $p$ can also be
 negative in general with the sign corresponding to spatial
 orientation, but we will assume $p>0$ in this example. The
 Hamiltonian $H$ is actually a constraint in this case, but we will
 not discuss aspects of constrained systems in the geometric
 formulation here.

 To simplify calculations we already weaken the notion of a strong
 effective system and require agreement between quantum and effective
 dynamics only up to corrections of the order $\hbar$. Performing the
 $\hbar$ expansion of the mean value of the Hamiltonian we obtain
\begin{equation}
 H_Q=H+\frac{1}{2}\hbar\kappa H,_{ij}\tilde{G}^{ij}+O(\hbar^{\frac{3}{2}})=
 H-\frac{3\hbar}{\gamma^2}\left(\sqrt{p}\tilde{G}^{0,2}+
\frac{c}{\sqrt{p}}\tilde{G}^{1,2}-\frac{c^2}{8\sqrt{p^3}}
 \tilde{G}^{2,2}\right)+O(\hbar^{\frac{3}{2}})\,.
\end{equation}
in terms of $\tilde G^{a,n}=\ell_{\rm P}^{-n}{G}^{a,n}$. These
variables are motivated by the uncertainty relations, with for the
symplectic structure in this example read
$G^{0,2}G^{2,2}-(G^{1,2})^2\geq \frac{1}{36}\gamma^2\ell_{\rm
P}^4$. Thus, one can expect that for minimal uncertainty the
$\tilde{G}$ (which are not dimensionless) do not contribute further
factors of $\hbar$. We will now perform a more detailed analysis.

{}From the commutation relation $[c,p]=\frac{1}{3}i\gamma\ell_{\rm
P}^2$ we obtain
\[
 \dot G^{a,n}=(\dot c\partial_c+\dot p\partial_p)G^{a,n}=
 -\frac{1}{\gamma\sqrt{p^3}}
 \left(-2ap^2G^{a-1,n}+(n-2a)cpG^{a,n}-\frac{(n-a)c^2}{4}G^{a+1,n}\right)\,.
\]
At this point it is useful to define $G^{a,n}=:c^{n-a}p^ag^{a,n}$ with
dimensionless $g$, leading to
\[
 ({\textstyle\frac{1}{2}}c\partial_c-2p\partial_p)g^{a,n}=-ag^{a-1,
 n}+\frac{1}{4}(n+a)g^{a,n}-\frac{1}{8}(n-a)g^{a+1,n}\,.
\]
This system of partial differential equations can be simplified by
introducing coordinates $(x,y)$ by $e^{2x}=\ell c^2/\sqrt{p}$ and
$y:=c^2\sqrt{p}/\ell$ with a constant $\ell$ of dimension length,
e.g.\ $\ell=\kappa E$ as the only classically available length scale
independent of the canonical variables, such that
$\frac{1}{2}c\partial_c-2p\partial_p=\partial_x$ and
$(\frac{1}{2}c\partial_c-2p\partial_p)f(y)=0$ for any function $f$
independent of $x$.

The general solution for $n=2$ then is
\begin{eqnarray}
g^{0,2}&=&g_0(y)+g_{\frac{3}{2}}(y)e^{\frac{3}{2}x}+g_{3}(y)e^{3x}\nonumber\\
g^{1,2}&=&2g_0(y)-g_{\frac{3}{2}}(y)e^{\frac{3}{2}x}-4g_{3}(y)e^{3x}\nonumber\\
g^{2,2}&=&4g_0(y)-8g_{\frac{3}{2}}(y)e^{\frac{3}{2}x}+16g_{3}(y)e^{3x}\nonumber
\end{eqnarray}
subject to the uncertainty relation
\begin{equation}\label{QCuncert}
4g_0g_3-g_{\frac{3}{2}}^2\geq
\frac{\gamma^2\ell_{\rm P}^4}{2^23^4\ell^{\frac{3}{2}}(c^2\sqrt{p})^{\frac
{5}{2}}}\,.
\end{equation}
Since $H$ is a constraint, $y$ will be constant physically such that
we can also consider $g_0$, $g_{\frac{3}{2}}$ and $g_{3}$ as
constants. On the constraint surface, the right hand side of the
uncertainty relation is then of the order $(\ell_{\rm P}/\kappa E)^4$
for the above choice of $\ell$ and thus very small.

Note first that, unlike the free particle and the harmonic
oscillator examples, solutions for the $G^{a,n}$ do not leave
unaffected the effective system. In this example, provided that it
allows an effective Hamiltonian description, we would thus encounter
an effective Hamiltonian different from the classical one. Spreading
back-reacts on the dynamics according to the effective equations
\begin{eqnarray}
 \gamma\dot c &=&-c^2p^{-\frac{1}{2}}\left(1+
\frac{1}{2}g_0-g_{\frac{3}{2}}(\ell c^2p^{-\frac{1}{2}})^{3/4}
 +11g_3(\ell c^2p^{-\frac{1}{2}})^{3/2}
+\cdots\right)
\\\gamma\dot p&=&c\sqrt{p}\left(4+
2g_0+
2g_{\frac{3}{2}}(\ell c^2p^{-\frac{1}{2}})^{3/4}
 -16g_3(\ell c^2p^{-\frac{1}{2}})^{3/2}
+\cdots\right)
\end{eqnarray}
There is no explicit $\hbar$ in the correction terms because we use
dimensionless variables, but the uncertainty relation shows that for
constants close to minimal uncertainty the corrections are of higher
order in the Planck length.

Moreover, as in the free particle case no constant solutions for the
$G^{a,n}$ exist.  We thus have to weaken not only the condition of a
preserved embedding, but also its horizontality. Since we are
interested in effective equations only up to a certain order in
$\hbar$, which we already used in the dynamics of this example, it is
reasonable to require constant $G$ also only up to terms of some
order in $\hbar$. This means that the quantum variables do not need to
be strictly constant, but change only slowly. In this example, we have
\begin{eqnarray*}
 \dot{G}^{0,2} &=& -\gamma^{-1}
 c^3 p^{-1/2}
(g_0+{\textstyle\frac{5}{2}}g_{\frac{3}{2}}e^{\frac{3}{2}x}+
4g_3e^{3x})\\
 \dot{G}^{1,2} &=& 3\gamma^{-1}
 c^2 p^{1/2}
(g_0+ 2g_3e^{3x})\\
 \dot{G}^{2,2} &=& 4\gamma^{-1}
 c p^{3/2}
(2g_0+5g_{\frac{3}{2}}e^{\frac{3}{2}x}+ 2g_3e^{3x})
\end{eqnarray*}
where $e^x$ is small for a large, semiclassical universe and the
dominant terms are given by $g_0$. For large $p$, $\dot{G}^{2,2}$
grows most strongly, but we can ensure that it is small by using small
$g_0$. It is easy to see that the uncertainty relation allows $g_0$ to
be small enough such that the $\dot{G}^{a,2}$ are small and at most of
the order $\hbar$. For instance, $g_{\frac{3}{2}}=0$, $g_3=1$ and
$g_0\sim \ell_{\rm P}^4 \ell^{-3/2} (c^2\sqrt{p})^{-5/2}$ is a
suitable choice where correction terms to the classical equations are
small and the strongest growth of the second order quantum variables,
given by $\dot{G}^{2,2}\sim \ell_{\rm P}^4 \ell^{-3/2} c^{-4}p^{1/4}$
is small on the constraint surface and using $\ell\sim \kappa E$:
$\dot{G}^{2,2}\sim \ell_{\rm P}^4 (\kappa E)^{-7/2} p^{5/4}$.  To the
$\hbar$-order of the equations derived here the system is thus almost
preserved, and quantum variables do not grow strongly for some time of
the evolution provided that the integration constants $g_a$ are chosen
appropriately. (Similar results, without using explicit quantum
variables $G$, have been obtained in \cite{Josh,Perturb}.)
\end{excount}

In the following section we will
formalize the weakened conditions on an effective system and show
that this allows one to reproduce standard effective action results.

\section{An-Harmonic oscillator}

We now come to the main part of this paper. As motivated by the
preceding examples, we first weaken the effective equation scheme
developed so far and then show that it reproduces the standard
effective action results when quantum dynamics is expanded around the
ground state of a harmonic oscillator. From what we discussed so far,
one can already see that basic properties are the same: First, the
harmonic oscillator ground state (or any coherent state) gives a
quantum dynamics with constant quantum variables such that the quantum
Hamiltonian differs from the classical one only by a
constant. Effective equations of motion are then identical to the
classical ones, which agrees with the usual result. If there is an
anharmonic contribution to the potential, however, the evolution of
classical variables depends on the quantum variables, and moreover
there is no finite set of decoupled quantum variables. Thus, for an
exact solution all infinitely many quantum variables have to be taken
into account, and in general {\em no strong effective system
exists}. This is the analog of the non-locality of the standard
effective action which in general cannot be written as a time integral
of a functional of the $q^i$ and finitely many of their time
derivatives. In standard effective actions, a derivative expansion is
an important approximation, and similarly we have to weaken our
definition of effective systems by introducing approximate notions.

The classical Hamiltonian is now given by $H=
\frac{1}{2m}p^2+\frac{1}{2}m\omega^2q^2+U(q)$, and the quantum
Hamiltonian in terms of dimensionless quantum variables (\ref{Gtilde}),
dropping the tilde from now on, is
\begin{equation} \label{HQexpand}
H_Q=\frac{1}{2m}p^2+\frac{1}{2}m\omega^2q^2+U(q)
+\frac{\hbar\omega}{2}(G^{0,2}+G^{2,2})+\sum_n\frac{1}{n!}
(\hbar/m\omega)^{n/2}U^{(n)}(q)G^{0,n}\,.
\end{equation}
This generates equations of motion
\begin{eqnarray}
\nonumber  \dot{q}&=& m^{-1}p\label{eom}\\
\dot{p}&=&-m\omega^2q -U'(q)-\sum_n\frac{1}{n!}(m^{-1}\omega^{-1}
\hbar)^{n/2}U^{(n+1)}(q)G^{0,n}\\
\nonumber \dot{G}^{a,n}&=&-a\omega
G^{a-1,n}+(n-a)\omega G^{a+1,n}
-\frac{aU''}{m\omega}G^{a-1,n}\\
\nonumber&&+
\frac{\sqrt{\hbar}aU'''(q)}{2(m\omega)^{\frac{3}{2}}}G^{a-1,n-1}G^{0,2}
+\frac{\hbar aU''''(q)}{3!(m\omega)^2}G^{a-1,n-1}G^{0,3}\\
\nonumber&&
-\frac{a}{2}\left(
\frac{\sqrt{\hbar}U'''(q)}{(m\omega)^{\frac{3}{2}}}G^{a-1,n+1}+\frac{\hbar
U''''(q)}{3(m\omega)^2}G^{a-1,n+2}\right)\\
\nonumber&&
+\frac{a(a-1)(a-2)}{3\cdot2^3}\left(
\frac{\sqrt{\hbar}U'''(q)}{(m\omega)^{\frac{3}{2}}}G^{a-3,n-3}+\frac{\hbar
U''''(q)}{(m\omega)^2}G^{a-3,n-2}\right)+\cdots
\end{eqnarray}
showing explicitly that a potential of order higher than two makes
the equations of motion for the $G^{a,n}$ involve $G^{a,n+1}$,
$G^{a,n+2}$ and so on, therefore requiring one to solve an infinite
set of coupled non-linear equations. However, for semiclassical
dynamics the $G^{a,n}$ should be small as they are related to the
spreading of the wave function. This allows the implementation of a
perturbative expansion in $\hbar^{1/2}$ powers to solve the equations
for $G$, where the number of degrees of freedom involved to calculate
the equations of motion for the classical variables up to a given
order is finite.

We emphasize that corrections appear at half-integer powers in
$\hbar$, except for the linear order. This is in contrast to what is
often intuitively expected for quantum theories, where only
corrections in powers of $\hbar$ are supposed to appear. (Correction
terms of half-integer order do not appear only if the classical
Hamiltonian is even in all canonical variables.) However, this is much
more natural from a quantum gravity point of view where not $\hbar$
but the Planck length $\ell_{\rm P}=\sqrt{\kappa\hbar}$ is the basic
parameter, which is a fractional power of $\hbar$ (see the quantum
cosmology example).

To solve the equations, we expand $G^{a,n}=\sum_e
G^{a,n}_e\hbar^{e/2}$. If we want to find a solution up to $k$th order
we have to calculate the solutions to (\ref{eom}) for $G^{0,2}$ up to
the order $k-2$ and $G^{0,3}$ to the order $k-3$. At the same time,
these will be functions of the $G^{a,n}$ to all orders up to
$G^{a,3+2(k-3)-l}_l$ for all positive integer $l\leq 2k-3$.

\begin{ex}
For $U(q)=\frac{\delta}{4!} q^4$ we have equations of motion
\begin{eqnarray} \nonumber\dot
G^{a,n}_0&=&\!-a\omega G_0^{a-1,n}\!\! +\!\!(n\!-\!a)\omega
G_0^{a+1,n} \!\! -\!\!\frac{\delta
q^2a}{2m\omega}G_0^{a-1,n}\!\!\\
\nonumber\dot {G}^{a,n}_1&=&\!-a\omega G_1^{a-1,n}\!\!
+\!\!(n\!-\!a)\omega G_1^{a+1,n} \!\! -\!\!\frac{\delta
q^2a}{2m\omega} G_1^{a-1,n}\!\!+\!\!\frac{\delta
aq}{2(m\omega)^{\frac{3}{2}}} G_0^{0,2} G_0^{a-1,n-1}\\\nonumber&&
-\frac{\delta a
q}{2(m\omega)^{\frac{3}{2}}}\left(G_0^{a-1,n+1}-
\frac{(a-1)(a-2)}{12} G_0^{a-3,n-3}\right)\\
\nonumber\dot { G}^{a,n}_2&=&\!-a\omega G_2^{a-1,n}\!\!
+\!\!(n\!-\!a)\omega G_2^{a+1,n} \!\! -\!\!\frac{\delta
q^2a}{2m\omega}G_2^{a-1,n}\!\!+\!\!\frac{\delta
aq}{2(m\omega)^{\frac{3}{2}}}( G_1^{0,2} G_0^{a-1,n-1}+ G_0^{0,2}
G_1^{a-1,n-1})\\\nonumber&& -\frac{\delta a
q}{2(m\omega)^{\frac{3}{2}}}\left(G_1^{a-1,n+1}-\frac{(a-1)(a-2)}{12}
G_1^{a-3,n-3}\right)\\\nonumber&& +\frac{\delta a}{3!(m\omega)^2}
G_0^{0,3} G_0^{a-1,n-1}-\frac{\delta a q}{6(m\omega)^2}\left(
G_0^{a-1,n+2}-\frac{(a-1)(a-2)}{4(m\omega)^2}G_0^{a-3,n-2}\right)
\end{eqnarray}
up to second order.
\end{ex}

Now, in order to construct a strong effective theory of the system we
would again have to find a submanifold which is invariant under the
action of the Hamiltonian.  The only dynamics contained in our quantum
degrees of freedom then comes via the submanifold: $\dot
G^{a,n}=\dot{x}^i\partial_iG^{a,n}$, e.g.\ for a potential
$U(q)=\frac{\delta}{4!} q^4$
\begin{equation}\label{G=section}
\dot
G^{a,n}=\left(\frac{1}{m}p\partial_q-\left(m\omega^2q
+\frac{\delta}{3!}q^3 +\frac{\hbar\delta q}{2m\omega}G^{0,2}
+\frac{\hbar^{3/2}\delta
}{3!(m\omega)^{\frac{3}{2}}}G^{0,3}\right)\partial_p\right)G^{a,n}\,.
\end{equation}
It seems convenient to perform an expansion in $\delta$ in addition to
$\hbar$ in order to solve the system of equations.  However, solutions
of these equations, perturbative or exact, are in general not single
valued functions of the classical variables and therefore an exactly
preserved semiclassical submanifold does not exist. In fact, we have
\begin{lemma}
 Let $({\cal H}, \hat{H})$ be a quantum mechanical system such that
 $\hat{H}=\frac{1}{2m}\hat{p}^2+V(\hat{q})$. If
 $({\cal H},\hat{H})$ admits a strong effective system of the
 classical dimension then
 $({\cal H}, \hat{H})$ is linear.
\end{lemma}
\begin{proof}
 By assumption, we have an embedding of the classical phase space into
 the quantum phase space such that the quantum flow is everywhere
 tangential to the embedding and the classical symplectic structure is
 induced. We can thus take the quantum Hamiltonian vector field and
 choose additional horizontal vector fields generated by functions
 $L^i$ on ${\cal K}$ such that they span the tangent space to ${\cal
 P}$ in each point $p\in{\cal P}$. Since, by construction, the
 collection of all those vector fields can be integrated to a
 manifold, they are in involution.  Vector fields on the bundle,
 finally, correspond to linear operators on the Hilbert space having
 the same commutation relations as the Poisson relations of the
 generating functions. There is thus a complete set of operators of
 the quantum system which includes the Hamiltonian and is in
 involution.
\end{proof}

The notion of a strong effective system then does not allow enough
freedom to include many physically interesting systems. Indeed, the
dynamics of a strong effective system does not significantly differ
from the classical one:

\begin{theo} \label{StrongEffHam}
 For any strong effective system of classical dimension, $H_{\rm
 eff}=H+{\rm const}$ differs from the classical Hamiltonian only by a
 constant of order $\hbar$.
\end{theo}
\begin{proof}
 {}From the preceding lemma it follows that a strong effective system
 can exist only when the Hamiltonian is at most quadratic in the
 complete classical phase space functions $L^i$. In an expansion as in
 (\ref{HQexpand}) we then have only the linear order in $\hbar$
 containing $G^{a,2}$. Since by assumption the strong effective system
 is of the classical dimension, horizontality implies that the
 $G^{a,2}$ are constant. Thus, $H_Q-H=\hbar c$ with a constant $c$,
 and $H_Q$ directly gives the effective Hamiltonian.
\end{proof}

If quantum degrees of freedom are included in a strong effective
system of dimension higher than the classical one, they are then only
added onto the classical system without interactions, which is not of
much interest.  On the other hand, for effective equations one is not
necessarily interested in precisely describing whole orbits of the
system, for which single valued solutions $G(q,p)$ would be required,
but foremost in understanding the local behavior compared to the
classical one, i.e.\ modifications of time derivatives of the
classical variables. The conditions for a strong effective system,
however, are requirements on the whole set of orbits of the system.
Thus, as noted before, we have to weaken our definition of effective
systems. We first do so in a manner which focuses on the finite
dimensionality of classical systems but ignores more refined notions
of semiclassicality:

\begin{defi} \label{EffSys}
An {\em effective system of order $k$} for a quantum system $({\cal
H},\hat{H})$ is a dynamical system $({\cal M},X_{\rm eff})$, i.e.\ a
finite-dimensional manifold ${\cal M}$ together with an effective flow
defined by the vector field $X_{\rm eff}$, which can locally be
embedded in the K\"ahler manifold ${\cal K}$ associated with ${\cal
H}$ such that it is {\em almost preserved}: for any $p\in{\cal M}$
there is an embedding $\iota_p$ of a neighborhood of $p$ in ${\cal K}$
such that $X_H(p)-\iota_{p*}X_{\rm eff}(p)$ is of the order $\hbar^{k+1}$
with the vector field $X_H$ generated by the quantum Hamiltonian.
\end{defi}

An effective system in this sense allows one to describe a quantum
system by a set of finitely many equations of motion, as we
encountered it before in the examples. The only concept of
classicality is the finite dimensionality, while otherwise the quantum
variables included in the effective system can change rapidly and grow
large even if an initial state has small fluctuations. Moreover, the
finite dimensional space of an effective system is not required to be
of even dimension or, even if it is of even dimension, to be a
symplectic space. In general, it is only equipped locally with a
pre-symplectic form through the pull-back of $\Omega$ on ${\cal K}$. A
stronger notion, taking these issues into account, is

\begin{defi} \label{HamEffSys}
 A {\em Hamiltonian effective system $({\cal P},H_{\rm eff})$ of order
 $k$} for a quantum system $({\cal H},\hat{H})$ is a
 finite-dimensional subspace ${\cal P}$ of the K\"ahler
 manifold ${\cal K}$ associated with ${\cal H}$ which is 
\begin{enumerate}
\item {\em symplectic}, i.e.\ equipped with a symplectic structure
$\Omega_{\cal P}=\iota^*\Omega_{\cal K}+O(\hbar^{k+1})$ agreeing up to
order $\hbar^{k+1}$ with the pull back of the full symplectic
structure, and
\item {\em almost preserved and Hamiltonian}, i.e.\
there is a Hamiltonian vector field $X_{\rm eff}$ generated by the
{\em effective Hamiltonian} $H_{\rm eff}$ on ${\cal P}$ such that for
any $p\in{\cal P}$ the vector $X_H(p)-X_{\rm eff}(p)$ is of the order
$\hbar^{k+1}$ with the vector field $X_H$ generated by the quantum
Hamiltonian.
\end{enumerate}
\end{defi}

By using a symplectic subspace we ensure that the commutator algebra
of the quantum system, which determines the symplectic structure on
${\cal K}$, is reflected in the symplectic structure of the effective
system. Moreover, as in the previous definition the dynamics of the
effective system is close to the quantum dynamics.  Still, the
effective Hamiltonian is not directly related to the quantum
Hamiltonian: one generally expands the quantum Hamiltonian in powers
of $\hbar$, solves some of the equations of motion for $G^{a,n}$ and
reinserts solutions into the expansion. Nevertheless, to low orders in
$\hbar$ most fluctuations can be ignored and it is often possible to
work directly with the quantum Hamiltonian as the expectation value in
suitably peaked states. This is the case for effective equations of
quantum cosmology \cite{Inflation, Perturb, Josh} where this procedure
has been suggested first.

In this definition, we still do not include any reference to the
corresponding classical system. In general, its dynamics will not be
close to the effective dynamics, but there are usually regimes where
this can be ensured for at least some time starting with appropriate
initial states. Also the symplectic structure $\Omega_{\cal P}$ can
differ from the classical one. This is realized also for effective
actions such as (\ref{EffAc}), where the symplectic structure also
receives correction terms of the same order in $\hbar$ as the
Hamiltonian.  The effective and classical symplectic structures are
close if the embedding of ${\cal P}$ in ${\cal K}$ is ``almost
horizontal'' which can be formalized by requiring that for any $p\in
{\cal P}$ and $v\in{\rm Hor}_p^{\Omega}{\cal K}$ there is a $w\in
T_p{\cal P}$ such that $w-v\in T_p{\cal K}$ is of some appropriate
order in $\hbar$.

We do not make this definition of almost horizontality more precise
since it turns out not to be needed to reproduce usual effective
action results. Moreover, its practical implementation can be rather
complicated: The quantum cosmology example showed that the order to
which one can ensure almost horizontality is not directly related to
the order in $\hbar$ to which equations of motion are expanded. If one
has an almost horizontal embedding, ignored quantum degrees of freedom
remain almost constant such that they do not much influence the
evolution for an appropriately prepared initial state.  But not any
system can be approximated in this manner, and so the condition of
almost horizontality implies that for some systems only higher
dimensional Hamiltonian effective systems exist. In such a case there
are some quantum degrees of freedom which can by no means be ignored
for the effective dynamics. On the other hand, in such a case it may
be difficult to guarantee the existence of a symplectic
structure. This happens, for instance, if the $G^{a,2}$ change too
rapidly, but not higher $G$. One can then use a 5-dimensional
effective system with variables $(q,p,G^{0,2},G^{1,2},G^{2,2})$ which
can only be pre-symplectic and thus not Hamiltonian.  Alternatively,
one can drop the condition of almost horizontality, but then has to
accept a new (pre-)symplectic structure which is not necessarily
related to the classical one by only correction terms. These
constraints show that a discussion of quantum variables in
higher-dimensional effective systems can be complicated if one insists
on the presence of a canonical structure.  Moreover, computing the
symplectic structure on the K\"ahler space and its pull-back to the
effective manifold in an explicit manner is usually complicated (see,
however, Sec.~\ref{DynCohStates} for a brief discussion).

We thus present a final definition which does not require
an explicit form of the quantum symplectic structure but is sufficient
for the usual setting of effective actions:
\begin{defi}
 An {\em adiabatic effective system of order $(e,k)$} for a quantum
 system $({\cal H},\hat{H})$ is an effective system $({\cal M},X_{\rm
 eff})$ of order $k$ in the sense of Def.~\ref{EffSys} such that the
 local embeddings are given by solutions up to order $e$ in an
 adiabatic expansion of those quantum variables not included as variables of
 the effective system.
\end{defi}
Here, adiabaticity intuitively captures the physical property of a
weak influence of quantum degrees of freedom on the classical ones: in
the adiabatic approximation they change only slowly compared to the
classical variables. Provided that a semiclassical initial state is
chosen it is then guaranteed that the system remains semiclassical for
some time.

This viewpoint is still much more general than the usual definition of an
effective action, and it allows much more freedom by choosing
different finite-dimensional subspaces.  For an explicit derivation of
effective equations, of course, one has to find solutions
$G^{a,n}(x^i)$ as they appear in the quantum Hamiltonian, which
requires one to solve an infinite set of coupled differential
equations for infinitely many variables. Only in exceptional cases,
such as integrable systems, can this be done without
approximations. Moreover, general solutions for $G^{a,n}(x^i)$ contain
infinitely many constants of integration which then also appear in the
effective equations after inserting the $G^{a,n}(x^i)$. On the one
hand, this allows much more freedom in choosing the states, such as
squeezed or of non-minimal uncertainty, to perturb around. But it also
means that one needs criteria to fix the integration constants in
situations of interest. One such situation is that of

\begin{theo}
 A system with classical Hamiltonian
 $H=\frac{1}{2m}p^2+\frac{1}{2}m\omega^2q^2+U(q)$ admits an adiabatic
 effective system of order $(2,1)$ whose dynamics is governed by the
 effective action (\ref{EffAc}).
\end{theo}

\begin{proof}
 In order to find the subspace ${\cal P}$ and the dynamics on it we
 expand the quantum Hamiltonian in powers of $\hbar$ and solve the
 equations of motion for $G^{a,n}$ in an adiabatic approximation.

 The adiabatic approximation of slowly varying fields in the equations
 of motion is an expansion in a parameter $\lambda$ introduced for the
 sake of the calculation, but in the end set to
 $\lambda=1$. Derivatives with respect to time are scaled as
 $\frac{\md}{\md t}\rightarrow\lambda\frac{\md}{\md t}$ and, expanding
 $G^{a,n}=\sum_eG^{a,n}_e\lambda^e$, the equations of motion
\[
 \dot x^i\partial_iG^{a,n}=\{G^{a,n},H_Q\}_Q
\]
 imply
\[
 \dot x^i\partial_iG_{e-1}^{a,n}=\{G_e^{a,n},H_Q\}_Q\,.
\]
 In addition to the adiabatic approximation we also perform a
 semiclassical expansion in powers of $\hbar$.  In what follows, we
 will calculate the first order in $\hbar$ and go to
 second order in $\lambda$ for $G^{a,2}$.

 To zeroth order in $\lambda$ the equations to solve are
\[
 0=\{G^{a,n}_0,H_Q\}_Q=\omega\left((n-a)G_0^{a+1,n}-a
 \left(1+\frac{U''}{m\omega^2}\right)G_0^{a-1,n}\right)
\]
 with general solution
\[
 G_0^{a,n}=(^{n/2}_{a/2})(^n_a)^{-1}\left(
 1+\frac{U''}{m\omega^2}\right)^{a/2}G^{0,n}_0
\]
 for even $a$ and $n$, and $G_0^{a,n}=0$ whenever $a$ or $n$ are
 odd. This still leaves the value of $G_0^{0,n}$ free, which will be
 fixed shortly. To first order in $\lambda$,
\[
 (n-a)G_1^{a+1,n}-a\left(1+\frac{U''}{m\omega^2}\right)
 G_1^{a-1,n}=\frac{1}{\omega}\dot{G}_0^{a,n}
\]
 implies
\begin{lemma}
\[
 \sum_{a\:{\rm even}}(^{n/2}_{a/2})\left(1+\frac{U''}{m\omega^2}\right)^\frac{n-a}{2}\dot{G}_0^{a,n}=0
\]
\end{lemma}
\begin{proof}
{}From the equation above\[\sum_{a}(^{n/2}_{a/2})\left(1+\frac{U''}
{m\omega^2}\right)^\frac{n-a}{2}\!\!\dot{G}_0^{a,n}=\sum_{a}(^{n/2}_{a/2})\left(1+\frac{U''}
{m\omega^2}\right)^\frac{n-a}{2}\!\!\left(
 (n-a)G_1^{a+1,n}-a\left(1+\frac{U''}{m\omega^2}\right)
 G_1^{a-1,n}\right)\]
manipulating the first term of the right hand side expression we
shift $a\to a-2$ leaving the limits for $a$ unaffected in the
summation to obtain
\[\sum_{a}\frac{(n/2)!\left(1+\frac{U''}
{m\omega^2}\right)^\frac{n-a+2}{2}}{((a-2)/2)!((n-a+2)/2)!}
 (n-a+2)G_1^{a-1,n}=\sum_{a}(^{n/2}_{a/2})a\left(1+\frac{U''}{m\omega^2}\right)^\frac{n-a+2}{2}
 G_1^{a-1,n}\] which cancels then the second term to finish the
 proof.
\end{proof}
This imposes a constraint on $G_0^{0,n}$ solved by setting
$G^{0,n}_0=C_n(1+\frac{U''}{m\omega^2})^{-n/4}$.  The remaining
constants $C_n$ are fixed to $C_n=\frac{n!}{2^n(n/2)!}$ by requiring
that the limit $U\rightarrow 0$ reproduces the quantum variables of
coherent states of the free theory (\ref{freemoments}) or equivalently
by requiring the perturbative vacuum of the quantum theory to be
associated to the vacuum of the effective system.  Therefore,
\[
 G_0^{a,n}=\frac{(n-a)!a!}{2^n((n-a)/2)!(a/2)!}
 \left(1+\frac{U''}{m\omega^2}\right)^{\frac{2a-n}{4}}\,.
\]
 We will need only the $n=2$ corrections to first order in $\hbar$,
 and the solution to the first order equations becomes trivial:
 $G^{1,2}_1=\frac{1}{2\omega}\dot G^{0,2}_0$, the rest being zero. To
 second order we have
\[
 G^{2,2}_2-\left(1+\frac{U''}{m\omega^2}\right)
 G^{0,2}_2=\frac{1}{\omega} \dot{G}_1^{1,2}=\frac{1}{2\omega^2}\ddot G^{0,2}_0
\]
 again leaving free parameters in the general solution to be fixed by
the next, third order from which we obtain
\[
 \left(1+\frac{U''}{m\omega^2}\right)\dot
 G^{0,2}_2+\dot G^{2,2}_2=0
\]
as in the Lemma before. The previous two equations can be combined to
a first order differential equations for $G_2^{0,2}$ in terms of known
solutions at lower orders:
\[
 \dot{G}_2^{0,2}- \frac{\dot{G}_0^{0,2}}{G_0^{0,2}} G_2^{0,2}+
\frac{1}{\omega^2} (G_0^{0,2})^2 \dddot{G}^{0,2}_0=0\,.
\]
Its general solution is
\[
 G_2^{0,2}= \left( c-2\omega^{-2} (G_0^{0,2})^{3/2}
\frac{\md^2}{\md t^2}(G_0^{0,2})^{1/2}\right) G_0^{0,2}
\]
where the integration constant $c$ can be fixed to $c=0$ by requiring
the correct free limit $U=0$ (for which the original two differential
equations imply $G_2^{2,2}=-G_2^{0,2}=0$).  From this, the solution to
the system is
\[
 G_2^{0,2}=-\frac{2}{\omega^2} (G^{0,2}_0)^{\frac{5}{2}}
 \frac{\md^2}{\md t^2}(G^{0,2}_0)^{1/2}=\frac{\left(
 1+\frac{U''}{m\omega^2}\right)^{-\frac{7}{2}}}
 {4\omega^2}\left(\left(1+\frac{U''}{m\omega^2}\right)\frac{U'''\ddot
 q+U''''\dot q^2}{4m\omega^2}-5\left(\frac{U'''\dot
 q}{4m\omega^2}\right)^2\right)\,.
\]

Finally, putting our approximate expressions for the quantum variables
back into the equations of the classical variables (\ref{eom}), we obtain
\begin{eqnarray}
&&\nonumber\left(m+\frac{\lambda^2\hbar(U''')^2}{2^5m^2\omega^5\left(
 1+\frac{U''}{m\omega^2}\right)
 ^{\frac{5}{2}}}\right)\ddot q+\frac{\lambda^2\hbar\dot
 q^2\left(4m\omega^2U'''U''''\left(1+\frac{U''}{m\omega^2}\right)-
 5(U''')^3\right)}
 {2^7m^3\omega^7\left(1+\frac{U''}{m\omega^2}\right)^{\frac{7}{2}}}\\
&&+m\omega^2q+U'+\frac{\hbar
U'''}{4m\omega\left(1+\frac{U''}{m\omega^2}\right)^{\frac{1}{2}}}=0
\end{eqnarray}
as it also follows from the effective action (\ref{EffAc}) after setting
$\lambda=1$.
\end{proof}

The proof demonstrates the role of the harmonic oscillator ground
state and its importance for fixing constants in the effective
equations. If there is no distinguished state, effective equations
contain free parameters incorporating the freedom of choosing an
initial state in which the system is prepared.

The role of adiabaticity here is the same as in the derivative
expansion of low energy effective actions, but even for an an-harmonic
oscillator are the effective systems defined here more general: we are
not forced to expand around a vacuum state but can make other choices
depending on the physical situation at hand. The vacuum state was used
here in order to fix the constants $C_n$ which appear when integrating
equations of motion for quantum variables. One can just as well choose
different constants, for instance those corresponding to a squeezed
state, and obtain the corresponding effective equations. Note,
however, that not every choice is consistent with the adiabatic
approximation. For instance, the proff showed that $G_0^{a,2}$ had to
be zero to leading order in $\hbar$. Thus, one cannot allow arbitrary
squeezing since the parameter $G_0^{1,2}$ is restricted. This can
become non-zero only at higher orders in the expansion.

Or, while one would always include the classical variables in the
effective system, they can be accompanied by some of the quantum
variables which are not treated as adiabatic. One can include such
quantum degrees of freedom directly as defined on the quantum phase
space, or introduce them by perturbing quantum variables around the
adiabatic solution, $G^{a,n}=G^{a,n}_{\rm adiabatic}+g^{a,n}(t)$. New
degrees of freedom given by $g^{a,n}(t)$ are then independent of the
classical variables and describe quantum corrections on top of the
adiabatic one.

There are also situations where no distinguished state such as the
vacuum is known, as it happens in the example of quantum cosmology
discussed earlier. General effective equations can then still be
formulated but contain free parameters incorporating the freedom of
choosing an initial state in which the system is prepared. The
constants $C_n$ in the above proof, for instance, would then remain
unspecified and appear in effective equations. To the same order as
considered here, only the constant $C_2$ enters which will appear in
general equations of motion. Following the lines of the proof above
without fixing $C_2$ is easily seen to lead to an effective action of
the form (\ref{EffAc}) with mass term
\[
 m+C_2^3\frac{\hbar U'''(q)^2}
 {2^5m^2\left(\omega^2+m^{-1}U''(q)\right)^{\frac{5}{2}}}
\]
and effective potential
\[
 -\frac{1}{2}m\omega^2q^2
-U(q)-C_2\frac{\hbar\omega}{2} \left(1+\frac{U''(q)}{m\omega^2}
\right)^{\frac{1}{2}}\,.
\]

\begin{rem}
 Knowing the effective action, one can derive the corresponding
 momentum and compute the effective symplectic structure. Corrections
 to the canonical symplectic structure can then occur if one uses a
 momentum variable $p$ that matches the dynamics of the mean value of
 $\hat p$.  Still, this does not necessarily imply that the system is
 a Hamiltonian effective system of first order as per
 Def.~\ref{HamEffSys} because we did not relate this symplectic
 structure to that following from pull-back from the quantum
 symplectic structure.
\end{rem}

\section{Dynamical coherent states}
\label{DynCohStates}

In addition to the effective dynamical behavior of classical and
quantum degrees of freedom it is also of interest to know
approximate states whose dynamics corresponds to the effective
evolution. Under the name of dynamical coherent states
\cite{CohStates}, they can be obtained by collecting the information
contained in the mean values of the fundamental operators and the
spreading as well as higher order distortions of the state of the
system. In this section, we only collect results related to the
previous discussion without going into further details.

As we already stated, the task could be achieved by summing
up the Hermite polynomial modes obtained through the Hamburger
momenta, but a short cut to the answer is possible using Moyal's
formula \cite{MoyalFormula} by which four arbitrary normalizable
vectors $|\Psi_1\rangle$, $|\Psi_2\rangle$, $|\Psi_3\rangle$ and
$|\Psi_4\rangle$ satisfy
\begin{equation}\label{moyal's}
 \int\frac{\md^2z}{2\pi}\langle\Psi_1|e^{z\hat
a^\dagger-\bar{z}\hat a}|\Psi_2\rangle \langle\Psi_3|e^{-z\hat
a^\dagger+\bar{z}\hat
a}|\Psi_4\rangle=\langle\Psi_1|\Psi_4
\rangle\langle\Psi_3|\Psi_2\rangle
\end{equation}
where $z=\frac{1}{\sqrt{2}}(z^q+iz^p)$ and $\hat a=
 \frac{1}{\sqrt{2\hbar}}(\hat q+i\hat p)$. For a bounded operator
$\hat F$, (\ref{moyal's}) can be rewritten as
\begin{equation}
  \int\frac{\md^2z}{2\pi}\langle\Psi_1|e^{z\hat
  a^\dagger-\bar{z}\hat a}|\Psi_2\rangle {\rm Tr}\left\{\hat F e^{-z\hat
  a^\dagger+\bar{z}\hat a}
\right\} =\langle\Psi_1|\hat
  F|\Psi_2\rangle\,.
\end{equation}

For given solutions $G^{a,n}$, the reconstruction of a dynamical
coherent state is completed by performing the integral with
arbitrary $|\Psi_1\rangle$, $|\Psi_2\rangle$ after inserting for
$\hat{F}$ the probability density operator $\hat\rho(q,p)$ and
assuming that the state is analytical such that
\[
  {\rm Tr}\left\{\hat\rho(q,p)e^{-z\hat a^\dagger+\bar{z}\hat a}
\right\}=e^{\frac{i}{\sqrt\hbar}(z^qp-z^pq)}
\sum_{n=0}^{\infty}\sum_{a=0}^{n}\frac{(-)^{n-a}i^n}{n!}(^n_a)
(z^q)^{a}(z^p)^{n-a}G^{a,n}(q,p)
\] 
produce the matrix elements of $\hat\rho(q,p)$ in a basis of
operators $e^{z\hat a^\dagger-\bar{z}\hat a}$. For the anharmonic
oscillator to $0th$ order in $\hbar$ we have $G^{i_1,\ldots,
i_n}=\frac{n!}{(n/2)!}G^{(i_1i_2}\cdots
G^{i_{n-1}i_n)}$ for $n$ even, implying
\[
  {\rm Tr}\left\{\hat
\rho_{_U}\!(q,p)e^{-z\hat a^\dagger+\bar{z}\hat a}
\right\}=\exp\left(\frac{i}{\sqrt\hbar}(z^i\epsilon_{ij}x^j)-\frac{1}{2}z^{i
}z^{j}\epsilon_{ik}\epsilon_{jl} G^{kl}(q,p)\right)\,.
\]
In order to perform the integral above, we choose to work with
Harmonic oscillator coherent states
$|\alpha\rangle=e^{\alpha\hat{a}^\dagger-\bar\alpha\hat{a}}|0\rangle$
for which the matrix elements of the exponential
operator are $\langle\alpha|e^{z\hat
a^\dagger-\bar{z}\hat{a}}|\alpha'\rangle=\exp(-\frac{1}{4\hbar}
({\alpha'}^i-\alpha^i)
\delta_{ij}({\alpha'}^j-\alpha^j)+\frac{i}{4\hbar}
({\alpha'}^i+\alpha^i)\epsilon_{ij}
({\alpha'}^j-\alpha^j+2z^j))$. Finally, defining
$S_i=\delta_{ij}({\alpha'}^j-\alpha^j)+
i\epsilon_{ij}({\alpha'}^i+\alpha^i-2x^i),$ the matrix elements of
the probability density operator are
\begin{eqnarray}\label{rho}
  \langle
\alpha|\hat\rho_{_U}\!(q,p)|\alpha'
\rangle &=& \frac{1}{\sqrt{\det(\frac{1}{2}\delta^{ij}+G^{ij})}}
\exp\left(-\frac{1}{4\hbar}S_{i_1}\epsilon^{i_1j_1}(2G^{ij}+
\delta^{ij})^{-1}_{j_1j_2}\epsilon^{j_2i_2}S_{i_2}
\right)\\
&&\times \exp\left(-\frac{i}{4\hbar}
({\alpha'}^i-\alpha^i)\epsilon_{ij}({\alpha'}^j+\alpha^j)
-\frac{1}{4\hbar}({\alpha'}^i-\alpha^i)\delta_{ij}({\alpha'}^j-\alpha^j)
\right).\nonumber
\end{eqnarray}

The trace of the operator above can now be computed to equal one
whenever $G^{ij}$ is a non-degenerate matrix. In order to be sure that
$\rho$ is a density matrix, we need to show its positivity. We do not
have a complete proof for arbitrary systems, but using the fact that
the assumption of the state being semi-classical requires the mean
values of operators to be given by their classical expressions up to
$\hbar$ corrections, a case by case study leads to the conclusion that
the positive mean values above lead to positivity of the
operator.

Furthermore, the state of the quantum system as given above is not in
general a pure state, but if
$G^{ij}=\frac{\hbar}{2}(e^{g\epsilon})^i_k(e^{g\epsilon})^j_l\delta^{kl},$
also $\hat\rho_{_U}(x)^2$ has trace one and thus gives a pure state which
can be realized as a squeezed coherent state labeled by the symmetric
matrix $g_{ij}$ through
\begin{equation}|x,g\rangle=\exp\left(\frac{i}{2\hbar}g_{ij}(\hat x^i-x^i)(\hat
x^j-x^j)\right)
\exp\left(-\frac{i}{\hbar}x^i\epsilon_{ij}\hat x^j\right)|0\rangle.
\end{equation}
With the help of $e^{-\frac{i}{2\hbar}g_{ij}\hat x^i\hat x^j}\hat
x^ke^{\frac{i}{2\hbar}g_{ij}\hat x^i\hat x^j}=
(e^{g\epsilon})^k_l\hat x^l$, the remaining fiber coordinates become
\begin{equation}
G^{i_1,\ldots, i_n}(g_{ij})=\frac{\hbar^{n/2}n!}
{2^{n}(n/2)!}(e^{g\epsilon})^{i_1}_{j_1}
\cdots(e^{g\epsilon})^{i_n}_{j_n}\delta^{(j_1j_2}\cdots
\delta^{j_{n-1}j_n)}
\end{equation}

Reconstructing a dynamical coherent state from the quantum variables
$G^{a,n}$ also provides means to compute the symplectic structure on
the effective space, as needed for a Hamiltonian effective system as
per Def.~\ref{HamEffSys}.  For the evaluation of the symplectic
structure on the vector fields we obtain the pull-back $\Omega(Y,Z)=
2\hbar{\rm Im}\langle Y,Z\rangle$ where $Y$ and $Z$
are tangent vectors to the embedded effective manifold. Given a
dynamical coherent state $|\psi(f^i)\rangle$ as a function of
classical variables $f^i$, we can define a basis of the tangent space
spanned by $|i\rangle:=\partial{|\psi\rangle}/\partial f^i$. Expanding
$Y=\sum_i Y_i|i\rangle$ and $Z=\sum_i Z_i|i\rangle$, we have
\[
  \langle Y|Z\rangle= \sum_{i,j} \bar{Y}^iZ^j 
\frac{\partial\langle\psi|}{\partial
f^i}\frac{\partial|\psi\rangle}{\partial f^j}
\]
such that we can formally write
\begin{equation}
\Omega
=-2i\hbar\,\md(\langle x_1,\ldots,x_n|)\wedge
\md(|x_1,\ldots,x_n\rangle)\,.
\end{equation}
Thus, the pull-back of the symplectic structure to the subspace of
squeezed states is
\begin{equation} \label{OmegaSqueezed}
\Omega_{|x,g\rangle}= 2\epsilon_{ij}\md x^i\wedge \md
x^j+2^{-5}\hbar \delta^{i_1i_2} \epsilon^{i_3i_4}
(\delta^{j_1}_{i_1}+ (e^{g\epsilon})^{j_1}_{i_1})\cdots
(\delta^{j_4}_{i_4}+ (e^{g\epsilon})^{j_4}_{i_4})\md
g_{j_1j_3}\wedge\md g_{j_2j_4}\,.
\end{equation} 
For an effective system of the classical dimension, corresponding to a
set of solutions $g_{ij}(x^k)$, we can further pull back
(\ref{OmegaSqueezed}) to the classical manifold and obtain the quantum
symplectic structure there. This shows that the classical symplectic
structure is reproduced up to corrections of order $\hbar$ if the $g$
do not change strongly (adiabaticity or almost horizontality), and
provides means to compute those correction terms.

\section{Conclusions}

Comparison with common effective action techniques applicable to
an-harmonic oscillators demonstrates how effective systems can be
formulated more generally for any quantum system. We have extracted
several definitions which have different strengths and use different
mathematical structures:
\[
 \includegraphics[width=6.0cm]{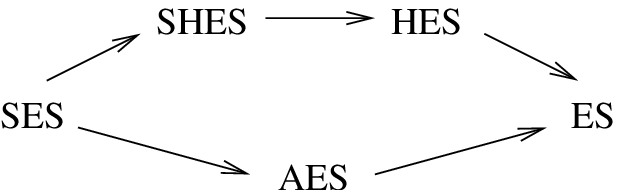}
\]
Here, the strengths of each of our
definitions are compared in a condensed diagram by use of
implication arrows and abbreviations in which  the initial S holds
for strong, H for Hamiltonian, A for adiabatic and ES for effective
system. The only definition not provided before is that of a strong
Hamiltonian effective system which is a Hamiltonian effective system
which is {\em exactly} preserved and whose symplectic structure is
{\em exactly} the pull-back of the quantum symplectic structure. It is
clear from the discussions before that any strong effective system is
also strong Hamiltonian, and examples lead to the conjecture that the
converse is also true. Still, since we are not aware of a proof, we
include strong Hamiltonian effective systems in this diagram.

While the definition of Hamiltonian effective systems is most
geometrical, adiabatic effective systems turn out to be more
practical and are more directly related to path integral techniques.
The weakest notion of an effective system can be applied to any
system but does not incorporate many classical aspects except for
finite dimensionality for mechanical systems. As the examples
showed, in particular that of quantum cosmology, the general
definitions provided here are more widely applicable and also
present a more intuitive understanding of possible quantum degrees
of freedom. Moreover, they are always switched on perturbatively,
and no non-analyticity in perturbation parameters as with higher
derivative effective actions arises.

The expansion of the quantum Hamiltonian also showed that in general
half-integer powers of $\hbar$ have to be expected in correction terms
and not just integer powers as often stated. The only exception is the
first order in $\hbar^{1/2}$ which does not appear because the
expectation value of variables $G^1$ would be zero by
definition. Half-integer powers do not appear only if one has a system
with a Hamiltonian even in all canonical variables, such as an
an-harmonic oscillator with an even potential, as it often occurs in
quantum field theories. These observations are relevant for quantum
gravity phenomenology because an expansion in the Planck length
$\ell_{\rm P}=\sqrt{\kappa\hbar}$ naturally involves half-integer
powers in $\hbar$. From the perspective provided here one can expect
all integer powers of the Planck length except for the linear one.

Other advantages are that the effective equations have a geometrical
interpretation where only real variables, unlike $q(t)$ in the usual
definition, occur. We are dealing directly with equations of motion
displaying only the relevant degrees of freedom, which are
automatically provided with an interpretation as properties of the
wave function, and can directly deal with canonical formulations in
which the scheme indeed arises most naturally. The techniques are
general enough for arbitrary initial states and systems with unbounded
Hamiltonians, as demonstrated by our quantum cosmology example. The
infrared problem of (\ref{EffAc}) for $m\to0$ is seen to arise only in
the adiabatic approximation, but can easily be treated by using more
general notions of effectivity such as by including the spreading
parameters $G^{a,2}$ in a pre-symplectic effective system.

As discussed briefly in the preceding section, techniques introduced
here can also be used directly at the quantum level and not just for
effective semiclassical approximations. In this context, we have
presented only first steps, but this already shows that the techniques
can give information on dynamical coherent states. This will then also
have helpful implications for the effective equation scheme itself
from which such states arise, as they can give a handle on computing
the pull-back of the full symplectic structure.

\section*{Acknowledgements}

We thank Abhay Ashtekar for several discussions and suggestions in the
early stages of this work. We are grateful to Emil Akhmedov, Benjamin
Bahr, Oscar Castillo, H\'ector Hern\'andez, Mikolaj Korzynski, Angel
Mu\~noz, Hanno Sahlmann and Thomas Thiemann for fruitful discussions
on different aspects of this work.  MB is grateful to the Isaac Newton
Institute for Mathematical Sciences, Cambridge for hospitality during
the workshop ``Global Problems in Mathematical Relativity,'' where
this paper was completed, and thanks the organizers Piotr Chrusciel
and Helmut Friedrich for an invitation.

\end{document}